\documentclass[aps,pre,superscriptaddress,twocolumn,eprint,amsmath,amssymb,longbibliography,nofootinbib,groupedaddress]{revtex4-1}

\usepackage{amsmath,amsthm,amsfonts,amssymb,amscd}
\usepackage{fullpage}
\usepackage{lastpage}
\usepackage{graphicx}
\usepackage{hyperref}
\usepackage{wrapfig}
\usepackage{enumerate}
\usepackage{fancyhdr}
\usepackage{mathrsfs}
\usepackage{mathtools}
\usepackage{braket}
\usepackage{slashed}
\usepackage{bm}
\usepackage[dvipsnames]{xcolor}
\usepackage{float}
\usepackage[margin=0.55in]{geometry}
\usepackage[english]{babel}

\graphicspath{ {Images/} }
\usepackage{subcaption}
\usepackage{comment}

\usepackage{tikz}
\usetikzlibrary{calc}
\usetikzlibrary{decorations.markings}

\usepackage[section]{placeins} 

\newcommand{\p}[2]{\ensuremath{\frac{\partial #1}{\partial #2}}} 

\newcommand{\beq}{\begin{equation}}
\newcommand{\eeq}{\end{equation}}

\begin{document}

	\title{Defect Dynamics in Active Polar Fluids vs. Active Nematics}
	\author{Farzan Vafa}
	\email{fvafa@ucsb.edu}
	\affiliation{Department of Physics, University of California Santa Barbara, Santa Barbara, CA 93106, USA}

	\date{\today}
	\begin{abstract}
		
		Topological defects play a key role in two-dimensional active nematics, and a transient role in two-dimensional active polar fluids. In this paper, we study both the transient and long-time behavior of defects in two-dimensional active polar fluids in the limit of strong order and overdamped, compressible flow, and compare the defect dynamics with the corresponding active nematics model studied recently. One result is non-central interactions between defect pairs for active polar fluids, and by extending our analysis to allow orientation dynamics of defects, we find that the orientation of $+1$ defects, unlike that of $\pm 1/2$ defects in active nematics, is not locked to defect positions and relaxes to asters.  Moreover, using a scaling argument, we explain the transient feature of active polar defects and show that in the steady state, active polar fluids are either devoid of defects or consist of a single aster. We argue that for contractile (extensile) active nematic systems, $+1$ vortices (asters) should emerge as bound states of a pair of $+1/2$ defects, which has been recently observed.  Moreover, unlike the polar case, we show that for active nematics, a linear chain of equally spaced bound states of pairs of $+1/2$ defects can screen the activity term. A common feature in both models is the appearance of $+1$ defects (elementary in polar and composite in nematic) in the steady state.
		
\end{abstract}

\maketitle

\section{Introduction}

In the context of biological systems, topological defects are ubiquitous, where they have been associated with cell extrusion~\cite{saw2017topological,kawaguchi2017topological}, changes in cell density~\cite{Copenhagen2020topological} and morphogenetic processes~\cite{Maroudas-Sacks2020topological}, among others. Here we will study defects in the context of active systems, which are composed of self-propelled  active units that move and exert forces on their surrounding by consuming energy, either internal or external~\cite{marchetti2013hydrodynamics,Simha2002}. One class of active matter is active nematics, which consists of head-tail symmetric active units that tend to align, locally generating nematic (apolar) order ~\cite{ramaswamy2003active,doostmohammadi2018active}. For sufficiently large activity, there is a proliferation of topological defects in the nematic texture~\cite{giomi2013defect,thampi2013velocity,giomi2015geometry,doostmohammadi2017onset,doostmohammadi2018active}, and understanding of the dynamics of topological defects has been advanced by treating the defects as quasiparticles~\cite{keber2014topology,narayan2007long,giomi2013defect,pismen2013dynamics,shankar2018defect,Shankar2019hydro,vafa2020multidefect,Zhang2020dynamics}.

Another class of active matter is active polar fluids, which consists of  active polar units that tend to align, locally generating polar order~\cite{ramaswamy2010mechanics,marchetti2013hydrodynamics,Chate2020dry}. The phase diagram of active polar fluids has been extensively studied (for example,~ \cite{Kung2006hydrodynamics,Giomi2010sheared,Giomi2012polar,Gopinath2012dynamical,Gowrishankar2016nonequilibrium,chen2016mapping,Chate2020dry}), and defects have been observed in for example~\cite{Dombrowski2004self,Riedel2005a,Sokolov2007concentration,wensink2012meso,Schaller2013topological}. In contrast to active nematics, since active polar fluids have long range order~\cite{Toner1995long,Toner1998flocks}, defects are not spontaneously generated, and if generated due to boundary effect for example, the defects are expected to be transient~\cite{Husain2017emergent,Mahault2018self,Chate2020dry}. That being said, aspects of dynamics of defects in active polar fluids have been studied in~\cite{Kruse2004asters,Kruse2005generic,elgeti2011defect,Gopinath2012dynamical,Schaller2013topological,Gowrishankar2016nonequilibrium,Husain2017emergent}. Here we study transient dynamics of defects, and give another perspective why they are transient. Applying the same argument to active nematics uncovers a 1D chain of $+1$ defects which screens the activity.
 
In this paper, we study both the transient and long-time behavior of defects in two-dimensional active polar fluids in the limit of strong order and overdamped, compressible flow. As in~\cite{Zhang2020dynamics,vafa2020multidefect}, we consider an approximation for the global texture motivated from the passive case where the defects are widely separated and quasi-static, and use the variational principle to find defect dynamics within this ansatz. Here I shall follow the general approach of~\cite{vafa2020multidefect}. In contrast to previous work on the active nematics model~\cite{narayan2007long,sanchez2012spontaneous,giomi2013defect,pismen2013dynamics,vafa2020multidefect}, in this model we find that there are no active self-propulsion terms for the lowest charge ($\pm 1$) energy excitations. Also in contrast to~\cite{vafa2020multidefect}, we obtain interactions between two defects that are neither central nor perpendicular to a central force; they are generically non-central. By extending this ansatz to allow orientation dynamics of defects, we find that the orientation of $+1$ defects, unlike that of $\pm 1/2$ defects in active nematics~\cite{vafa2020multidefect}, is not locked to defect positions and relaxes to asters, which we confirm with simulations. Moreover, using a scaling argument, we explain the transient feature of active polar defects and show that in the steady state, active polar fluids are either devoid of defects or consist of a single aster. We argue that for contractile (extensile) active nematic systems, $+1$ vortices (asters) should emerge as bound states of a pair of $+1/2$ defects, which has been studied in~\cite{duclos2017topological,shankar2018defect,kumar2018tunable,Turiv2020topology,Thijssen2020,Pearce2020,Thijssen2020binding}. Moreover, unlike the polar case, we show that for active nematics, a linear chain of equally spaced bound states of two $+1/2$ defects can screen the activity term.  This hints at the existence of stationary lattice of bound states of pairs of $+1/2$ defects in the long term behavior of active nematics, perhaps similar to~\cite{oza2016antipolar,Thijssen2020}. A common feature in both models is the appearance of +1 defects (elementary in polar and composite in nematic) in the steady state.

The paper is organized as follows. We introduce the model in Sec.~\ref{sec:model} and in Sec.~\ref{sec:stationary} we review the class of quasi-stationary multi-defect solutions we use to parameterize the dynamics of textures. In Sec.~\ref{sec:results} we review the derivation of defect dynamics equations and present our results for the active induced pair-wise interactions. In Sec.~\ref{sec:pol} we extend our method to study orientation dynamics of defects, and in Sec.~\ref{sec:scaling} we offer an explanation as to why defects are transient and describe the long-time behavior. Finally, in Sec.~\ref{sec:comparison} we compare this model to the active nematics model introduced recently in~\cite{vafa2020multidefect}. Most of the technical details are relegated to Appendices~\ref{app:A}-\ref{app:pol}.

\section{The Model}
\label{sec:model}

We consider a two-dimensional polar fluid with density $\rho$ and vector order parameter $\bf p$ described by the free energy~\cite{de1993physics,Kung2006hydrodynamics} ${\mathcal F}(\{{\bf p}\})$:
\beq \mathcal F(\{{\bf p}\}) = \mathcal F_n(\{{\bf p}\}) + \mathcal F_p(\{{\bf p}\})\; ,\eeq
where
\beq {\mathcal F}_n(\{{\bf p}\})=\frac{1}{2}\int dxdy \left [C\left( \frac{\delta\rho}{\rho_0}\right)^2 + K  \ Tr ({ \nabla} {\bf p})^2+g(1- {\bf p}^2)^2 \right ]\;,
\eeq
\beq \mathcal F_p(\{{\bf p}\}) = \int dx dy B \frac{\rho}{\rho_0}\nabla\cdot\bf p\eeq
and $\rho_0$ is the equilibrium value of $\rho$.

The first term, ${\mathcal F}_n(\{{\bf p}\})$, is the usual free energy of a liquid crystal which contains only terms even in $\bf p$~\cite{de1993physics}, and the second term,  $\mathcal F_p(\{{\bf p}\})$, contains additional terms that break this $\bf p \to -\bf p$ symmetry. $K$ is the Frank constant in the one-constant approximation, and $g$ controls the strength of polar order. We assume to be deep in the ordered state ($g \rightarrow \infty$), where the coherence length $\xi = \sqrt{K/2g}$ is the smallest relevant lengthscale and $|\vec p|\approx 1$ except within polar defect cores of size $a \sim \xi$. Although symmetry allows us to write terms that are odd in $\bf p$ as in $\mathcal F_p$, and that density fluctuations are generally important for polar fluids, for simplicity of analysis and in order to connect with a nematic we will assume that this contribution due to $\mathcal F_p(\{{\bf p}\})$ can be ignored, for example by imposing $\bf p \to -\bf p$ symmetry, or assuming that $b$ is small, or we are in a region where gradients in density are small.

Relaxation towards the minimum of the free energy while advection by flow $\mathbf{v}$ leads to
\beq
\partial_t p_i+ {\bf v} \cdot \nabla {p}_i + \omega_{ij}p_j = - \frac{D}{4K}  \frac{\delta {\mathcal F}}{\delta {p}_{i} }\;,
\label{eq:p}
\eeq
where $D$  is the diffusivity and $\omega_{ij} =(\partial_i v_j-\partial_j v_i)/2$ is the vorticity. In the overdamped limit, $\mathbf v = v_0 \mathbf p$, where $v_0$ has the dimensions of a speed and represents the speed of an isolated active particle. With this assumption, our equations now take the form of the Toner-Tu equations~\cite{Toner1995long,Toner1998flocks,Toner2005hydrodynamics} (see~\cite{Souslov2017} for a clear exposition):
\beq
\partial_t p_i+ \frac{v_0}{2}{\bf p} \cdot \nabla {p}_i = - \frac{D}{4K}  \frac{\delta {\mathcal F}}{\delta {p}_{i}}\;,
\eeq

In Eq.~\eqref{eq:p} we have dropped the rate of strain alignment term~\cite{Kung2006hydrodynamics,ramaswamy2010mechanics} because in $2D$ and in the overdamped limit, its effect on dynamics can be represented by renormalizing the advection term. We rescale length with $\ell$, where $\ell$ is the characteristic separation between topological defects, and time with $\tau=\ell^2/D$. We assume that defects are widely separated, that is $\ell \gg \xi$, and thus define the dimensionless small parameter $\epsilon=\xi /\ell \ll 1$. We also define the dimensionless activity parameter $\lambda= v_0/8 D$.

As in~\cite{vafa2020multidefect}, it is convenient to adopt the language of complex analysis. In terms of complex coordinates $z=x + iy$ and ̄$\bar z = x - iy$, the complex partial derivatives $\partial = \partial_z= \frac{1}{2}(\partial_x-i\partial_y)$ and ${\bar \partial} =\partial_{\bar z}= \frac{1}{2} (\partial_x+i\partial_y)$, and the complex order parameter $p = p_x + ip_y$, 
the (dimensionless) free energy takes the form
\beq
{\mathcal F}(\{p\})= \int dzd{\bar z} \left [4|\partial p|^2+\epsilon^{-2}
(1- |p|^2)^2 \right ]\;.
\eeq
Finally, the equation of motion can be written as
\beq
\partial_t p = \mathcal I(p) = -\frac{\delta {\mathcal F}(\{p\})}{ \delta {\bar p}}+\lambda \mathcal I_\lambda(p)\;,
\label{eq:complexp}
\eeq
where
\beq \mathcal I_\lambda (p)=  -(p\partial + \bar p \bar \partial)p \; .\eeq

\section{Stationary and quasi-stationary  textures  deep in the ordered state}
\label{sec:stationary}

For simplicity, we first consider the passive case where $\lambda = 0$. Then we are interested in solving
\beq \partial_t p = -\frac{\delta {\mathcal F}(\{p\})}{\delta {\bar p}} = 4\partial \bar \partial p + 2\epsilon^{-2}(1 - |p|^2) \;.\eeq
Since this model was studied in~\cite{vafa2020multidefect}, we will simply review it here. The single defect solution is 

\beq
p=\psi (z, {\bar z})=A(|z|)\left (  \frac{z}{|z|} \right )^{\sigma}\;,
\eeq
with the amplitude $A(|z|)$ describing the defect core~\cite{Pismen1999}:  as $r\to0$, $A(r) \propto r$, and for $r \gg \epsilon$, $A(r) \simeq 1 - \frac{\epsilon^2}{4r^2}$ (see Appendix~\ref{app:A} for more details about $A$).

The multi-defect solution takes the form
\beq
p_0(z, {\bar z}|\{z_i\})= e^{i\psi}\prod_i \Psi_i = e^{i\psi}\prod _i A(|z - z_i|)\left (  \frac{z-z_i}{|\bar z - \bar z_i|} \right )^{\sigma_i} \;,
\eeq
where $\psi$ is the phase of $p$ at infinity. This texture satisfies the boundary condition $p \rightarrow e^{i\psi} e^{i\varphi \sum_i \sigma_i}$ as $|z| \rightarrow \infty$, where $\varphi$ is the polar angle. In the special case of a charge neutral system, $\sum_i \sigma_i = 0$, and so $p$ is constant on the boundary.
 
In the limit $\epsilon\to 0$, the multi-defect texture $p_0 (z,{\bar z}|\{ z_i \})$ is the minimizer of ${\mathcal F}(p)$ when defects are pinned (see e.g.~\cite{pacard2000linear} and references within). In terms of the defect positions $z_i$, the free energy ${\mathcal F}_0={\mathcal F}(p_0)$ takes the well-known form
\beq
{\mathcal F}_0 \approx 2 \pi \sum _{i \ne j} \sigma_i \sigma_j \log  \frac{|z_j-z_i|}{L},
\label{eq:F-Coulomb}
\eeq
which describes a Coulomb interaction between defect charges~\cite{chaikin2000principles}, where $L$ is the system size. Due to the Coulomb interaction, even in the absence of any ``activity", the defect cores will move to minimize the free energy ${\mathcal F}_0$. Thus even though $p_0$ textures minimize the free energy when defects are pinned, they are only quasi-static when the defects are no longer pinned. 

\begin{figure}[]
	\centering
	\subcaptionbox{$+1$ defect}
	{\includegraphics[width=0.45\columnwidth]{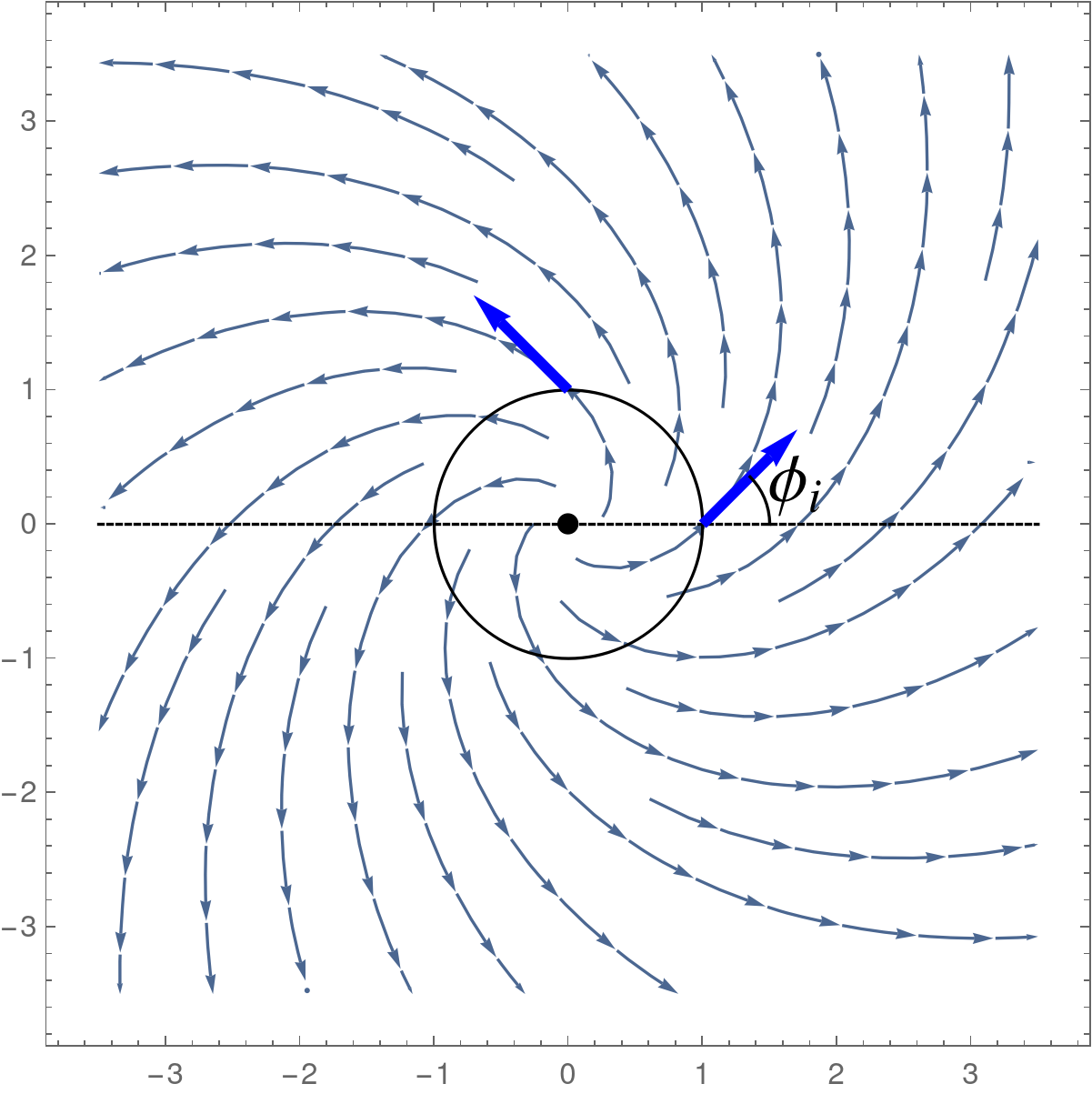}}
	\subcaptionbox{$-1$ defect}
	{\includegraphics[width=0.45\columnwidth]{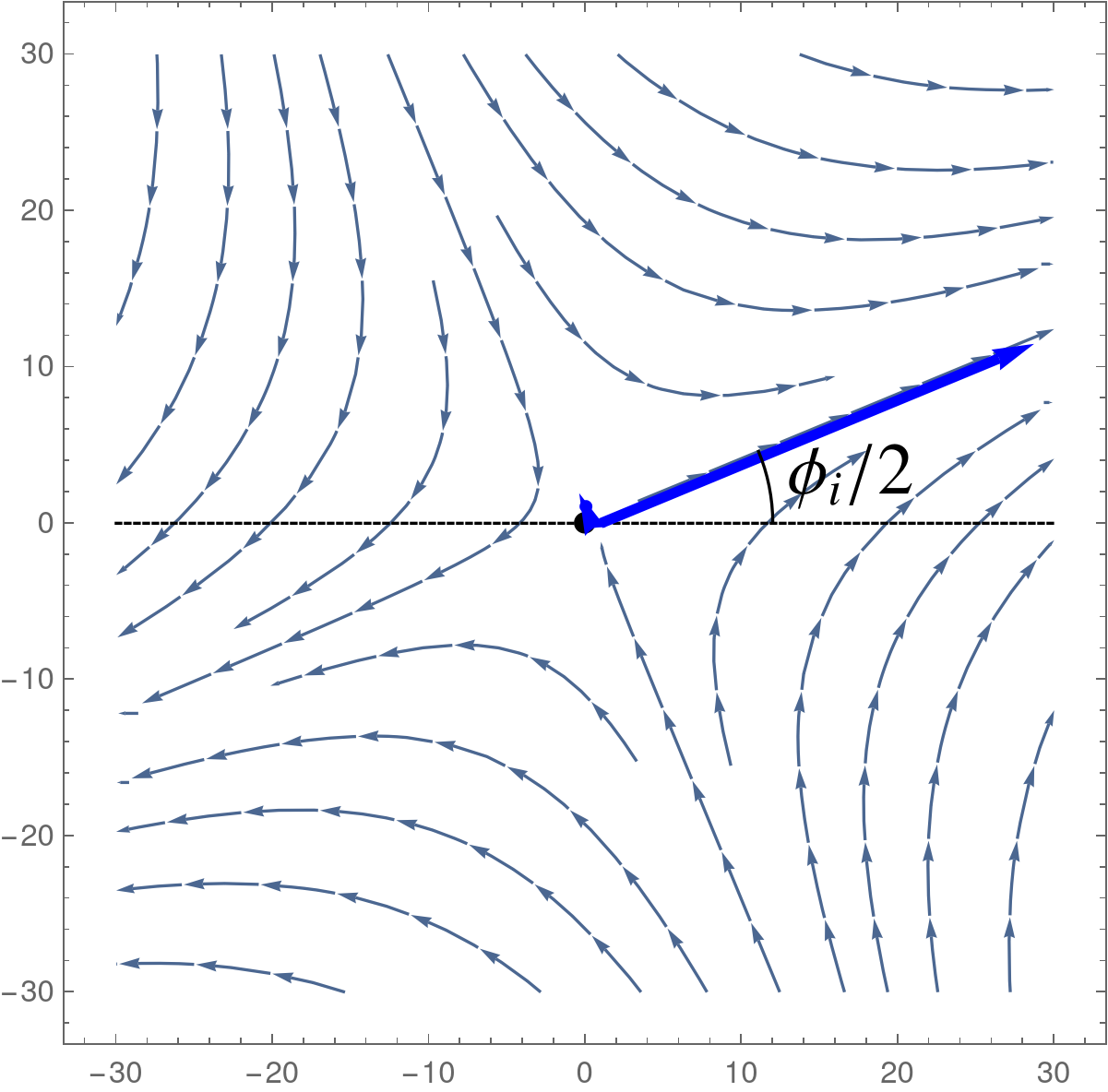}}\\
		\subcaptionbox{aster}
	{\includegraphics[width=0.45\columnwidth]{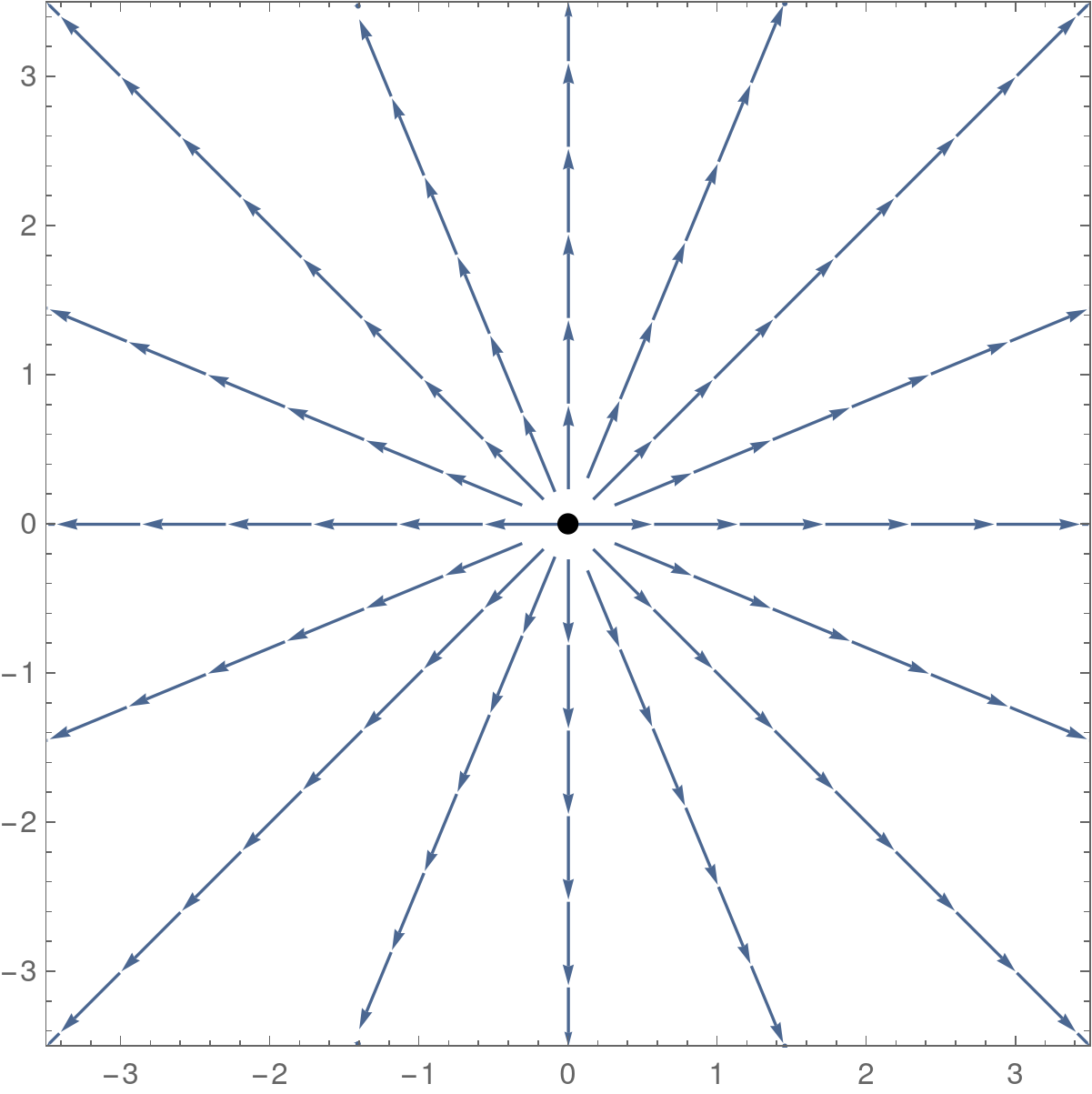}}
	\subcaptionbox{vortex}
	{\includegraphics[width=0.45\columnwidth]{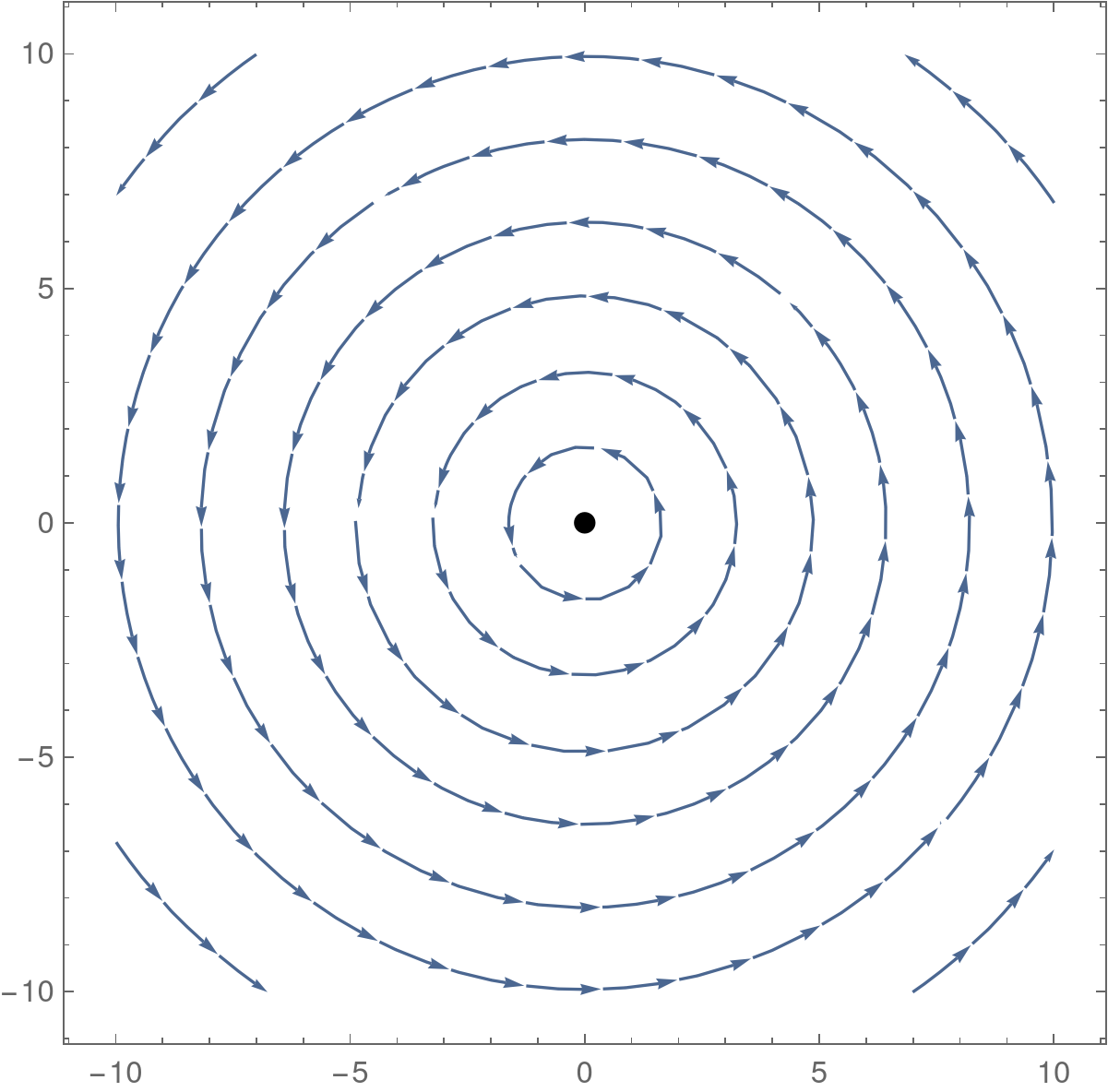}}
	\caption{Sketches of single defect textures showing the angle $\phi_i$ (the phase of $P_i$) for (a) a $+1$ defect where $\phi_i$ is the angle between $P_i$ and $\hat r$, (b) a $-1$ defect where $\phi_i/2$ is the angle of the separatrix. Special values of $\phi_i$ are shown in (c) and (d) for a $+1$ defect: (c) is an aster ($\phi_i=0$), and (d) is a vortex ($\phi_i=\pi/2$).}
	\label{fig:pols}
\end{figure}

As noted in~\cite{vafa2020multidefect}, near a defect $z_i$, we can write
\beq
p_0 (z, {\bar z}) \approx P_i \Psi_i (z-z_i, {\bar z}-{\bar z}_i)\;.
\eeq
where
\beq
P_i =  e^{i\phi_i} = e^{i\psi}\prod _{j \ne i} \left (  \frac{z_i-z_j}{|z- z_j|} \right )^{\sigma_j}   \;. \label{eq:phase}
\eeq
is a phase factor that will play an important role in the active induced dynamics of the defects. See Fig.~\ref{fig:pols} for a geometrical interpretation.

Finally, we note that for a global rotation, under which $z\to e^{i\eta}z$, the complex order parameter transforms as $p_0 \to p_0 e^{i\eta(1-\sum_i \sigma_i)}$. This implies that if $\sum_i \sigma_i \neq 1$, we can choose $\eta$ such that it eliminates the global phase factor $\psi$. In particular, we cannot eliminate the phase for a single $+1$ defect. This obstruction is not surprising since $+1$ defects are unique among defects in that they are rotationally invariant as $p \propto z$. We will see in our analysis that $\psi$ plays a crucial role for $+1$ defects.

\section{Dynamics of active polar defects (interactions)}
\label{sec:results}
\subsection{Method}

We are interested in solving the following PDE:
\beq \p{p}{t} = \mathcal I(p)\eeq
We do so by following the variational method used in~\cite{Zhang2020dynamics,vafa2020multidefect}, which we now review.  We start by making the ansatz
\beq p(z,\bar z,t) = p_0(z, \bar z,\{w_a(t)\})\eeq
where $w_a(t)$ (perhaps infinitely many) are parameters that need to be specified. (For example, $w_a(t)$ can include the defect positions, but is not strictly limited to them.) Once specified, $w_a(t)$ are computed by minimizing the deviation of $dp_0/dt$ from that described by the equation of motion, Eq.~\eqref{eq:complexp}. 
In other words, we minimize the error
\begin{align}
E &= \int d^2dz d\bar{z} \left| \partial_t p( z,{\bar z},t) - \frac{d}{dt} p_0( z,{\bar z}|\{ w_a (t) \})\right|^2 \nonumber\\
&\approx \int dz d\bar{z} \left| \mathcal I (p_0) - \dot w_a\p{p_0}{w_a} \right|^2
\label{eq:E}
\end{align}
with respect to $\dot w_a$, where $\mathcal I$ is defined in Eq.~\eqref{eq:complexp}. Of course, the goodness of our minimization depends on the ansatz and the chosen parameters $w_a$. We choose our ansatz to be $p_0$, because we know that when the defects are fixed and when $\lambda=0$, $p_0$ is a good solution~\cite{pacard2000linear}. Specifically, we assume that the defects are far away from each other and that $\lambda\ll 1$, in which case $p_0$ is a quasi-static solution to Eq.~\eqref{eq:complexp}. Taking into account that $\lambda\neq 0$ and the defects are not infinitely far away from each other leads to motion of the defects, and we will assume that the time-dependence of $p$ is only through the defect positions $z_i(t)$, and that the motion is slow. In other words, we will make the ansatz
\beq p(z,\bar z,t) = p_0(z, \bar z,\{z_i(t)\})\label{eq:ansatz}\eeq
where we have chosen $w_a(t)$ to be $z_i(t)$, the defect positions.

Doing so, one finds that~\cite{vafa2020multidefect}
\beq \mathcal M_{ij}\dot z_j  + \mathcal N_{ij}\dot{\bar z}_j   = -\p{{\mathcal F}_0}{\bar z_i} + \lambda{\mathcal U}_i\;, \label{eq:ODE}\eeq
where
\begin{align}
\mathcal M_{ij} &= \int d^2z[\bar\partial_i\bar p_0 \partial_j p_0 + \bar\partial_i p_0 \partial_j \bar p_0]\\
\mathcal N_{ij} &= \int d^2z[ \bar\partial_i\bar p_0 \bar \partial_j p_0 + \bar\partial_i p_0 \bar\partial_j \bar p_0]\;.
\end{align}
are the mobility matrices,
\beq \mathcal F_0 = -2\pi \sum_{i\not=j} \sigma_i\sigma_j \ln \frac{|z_i - z_j|}{L} \; , \eeq
is the Coulomb free energy, and
\beq
{\mathcal U}_i = \int d^2z[\bar\partial_i\bar p_0 \mathcal I_\lambda + \bar \partial_i p_0 \bar {\mathcal I}_\lambda]\label{eq:U_idef}\;.
\eeq
The mobility matrices $\mathcal M_{ij}$ and $\mathcal N_{ij}$ have been calculated in~\cite{vafa2020multidefect} to be
\begin{align}
\mathcal M_{ij} &\approx \pi \sigma_i\sigma_j \ln \frac{L}{r_{ij}} \\
\mathcal N_{ij} &\approx 0 \; .
\end{align}

Before proceeding, we would like to emphasize that in order to determine $z_i$, we are doing a global fit within our ansatz that finds the $z_i$ that minimizes the error. That is to say, although we interpret $z_i$ as the positions of defects, $z_i$ are simply parameters in our ansatz for the global texture that act as a proxy for the defect positions, and similarly $\dot z_i$ are not the true velocities of the defects. If we were interested in calculating the exact defect velocities, then we could do so with a local calculation which tracks the zeros of $p$. However, we are interested in how $p$ evolves everywhere, not just at specific points, which is why we minimize the error $E$ in Eq.~\eqref{eq:E}. Note that the fact that our equations depend on the system size $L$ is not surprising given we are doing a global fit in a region of size $L$. And, we have the freedom, if we are interested, to focus on the physics in a subregion of size $\ell < L$ by minimizing Eq.~\eqref{eq:E} in this subregion.

\subsection{Interactions}

\begin{figure}[]
	\centering
	\subcaptionbox{$-1$ and $+1$ defects}
	{\includegraphics[width=.45\columnwidth]{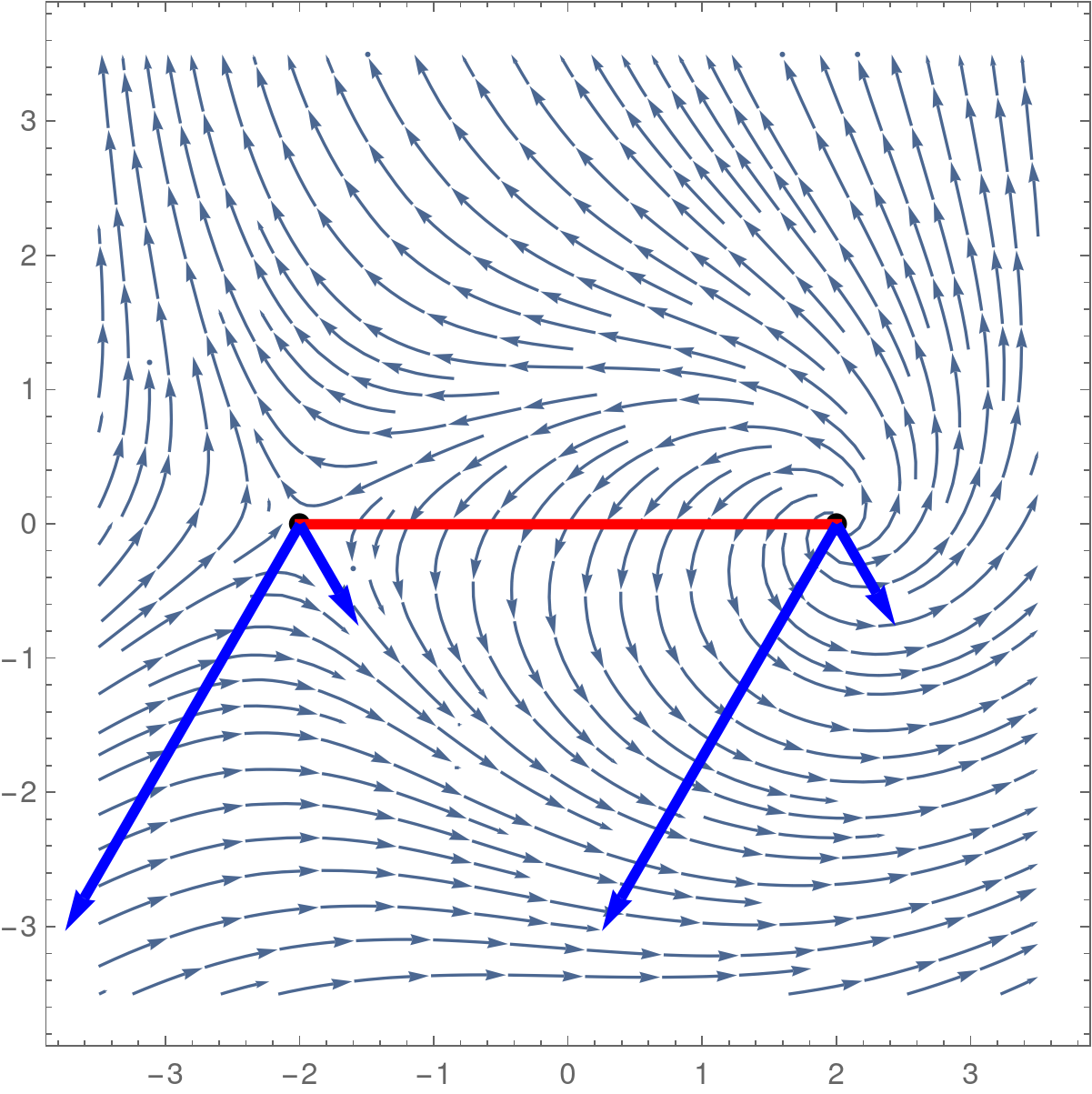}}
	\subcaptionbox{$-1$ and $-1$ defects}
	{\includegraphics[width=.45\columnwidth]{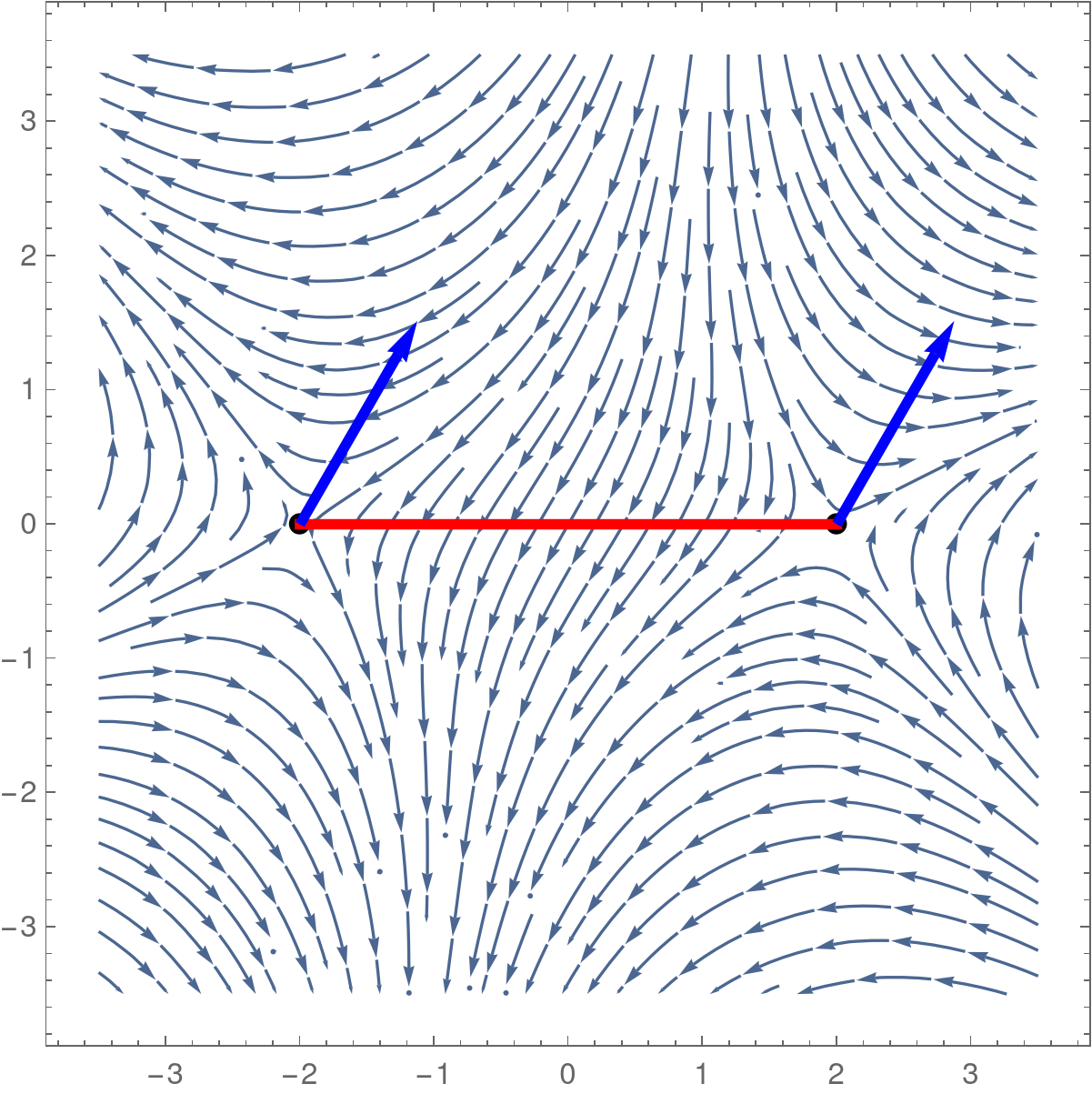}}
	\caption{Sketches of the active forces $f_{ij}$ for $\lambda>0$. The blue arrows denote the two components of the active force $f_{ij}$, and the red line joins the center of the two defects. For each pair, $f_{ij} = f_{ji}$, and the net forces are generically  non-central.
	}
	\label{fig:pair-forces}
\end{figure}

In Appendix~\ref{app:U_i}, we show that $\mathcal U_i$ (defined in Eq.~\eqref{eq:U_idef}) can be explicitly written in terms of the defect positions as 
\beq \lambda \mathcal U_i = -8\pi\ln \frac{L}{a}\lambda \bar P_i \delta_{\sigma_i,2} + \sum_{j\neq i} f_{ij},
\label{eq:U_i}\eeq
where in terms of the unit vector $\hat z_{ij} = (z_i - z_j)/|z_i - z_j|$ and its complex conjugate $\hat {\bar z}_{ij}$,
\beq f_{ij} = \frac{1}{2}\lambda\sigma_i\sigma_j \hat z_{ij} \left(P_i \hat z_{ij}^{\sigma_i -1}I^{(1)}_{ij} - \bar P_i \hat {\bar z}_{ij}^{\sigma_i -1}I^{(2)}_{ij}\right) \; \eeq
with
\begin{gather}
I^{(1)}_{++} = I^{(1)}_{--} = 2\pi \nonumber\\
I^{(1)}_{+-} = I^{(1)}_{-+} =  2\pi \ln \frac{L}{r_{ij}} + \mathcal O(L^0) \nonumber\\
I^{(2)}_{++} = 2\pi \ln \frac{L}{r_{ij}} + \mathcal O(L^0); \quad I^{(2)}_{--} = 0 \nonumber\\
I^{(2)}_{+-} = I^{(2)}_{-+} = 2\pi \; .
\end{gather}
The first term in Eq.~\eqref{eq:U_i} is the ``self-propulsion'' of a $+2$ defect along the $\bar P_i$ direction, where $P_i$ was defined in Eq.~\eqref{eq:phase}. Of course, we should not take this term too seriously, because a $+2$ defect can be interpreted as a bound state of two $+1$ defects, which is unstable because of the Coulomb repulsion. The second term in Eq~\eqref{eq:U_i} is the active induced pair-wise interaction, and its leading dependence on distance $r_{ij}$ between two defects $i$ and $j$ is $\ln L/r_{ij}$.

\begin{figure}[]
	\centering
	\subcaptionbox{Two $+1$ defects}
	{\includegraphics[width=0.49\columnwidth]{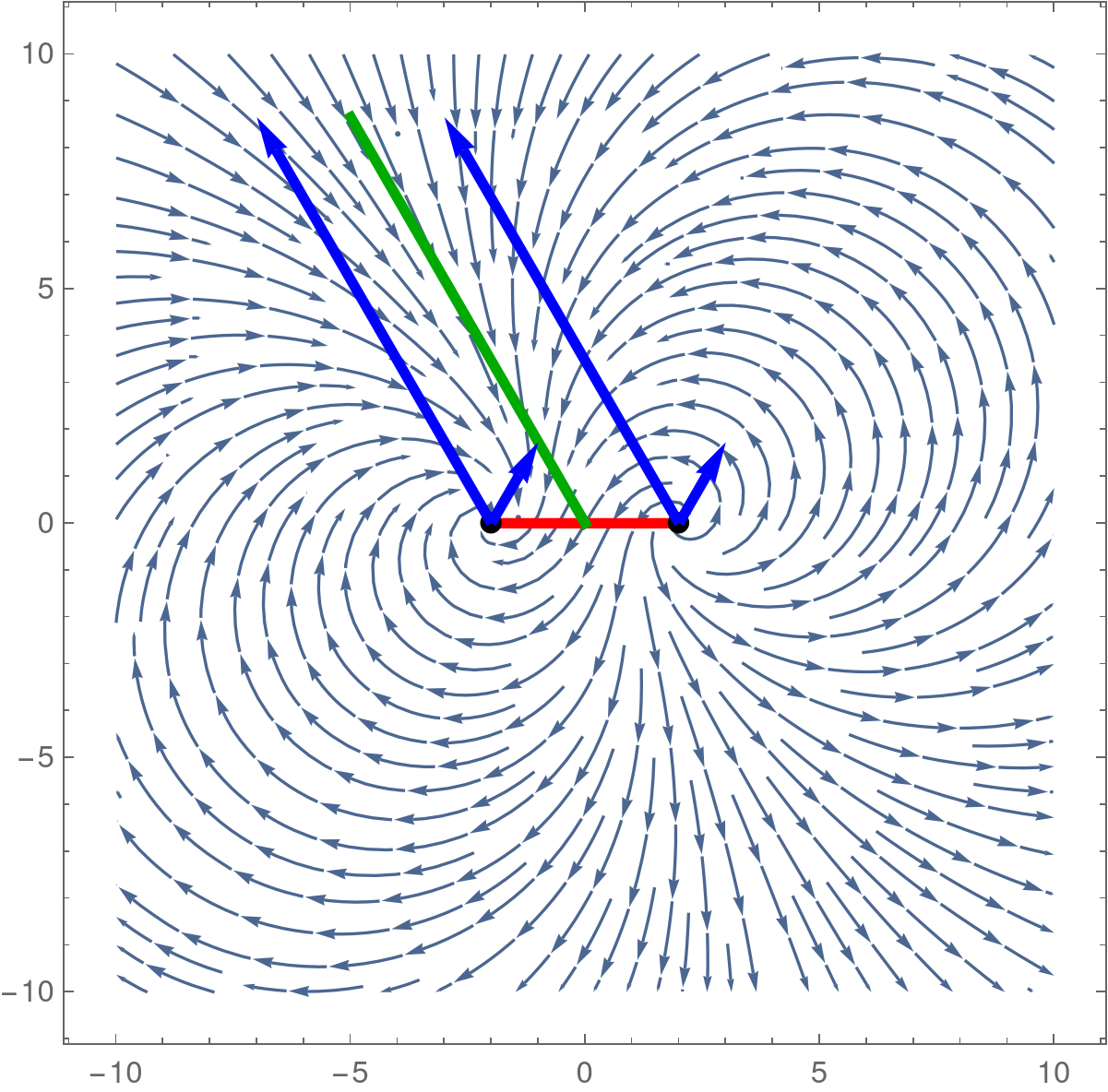}}
	\subcaptionbox{A single $+2$ defect}
	{\includegraphics[width=0.49\columnwidth]{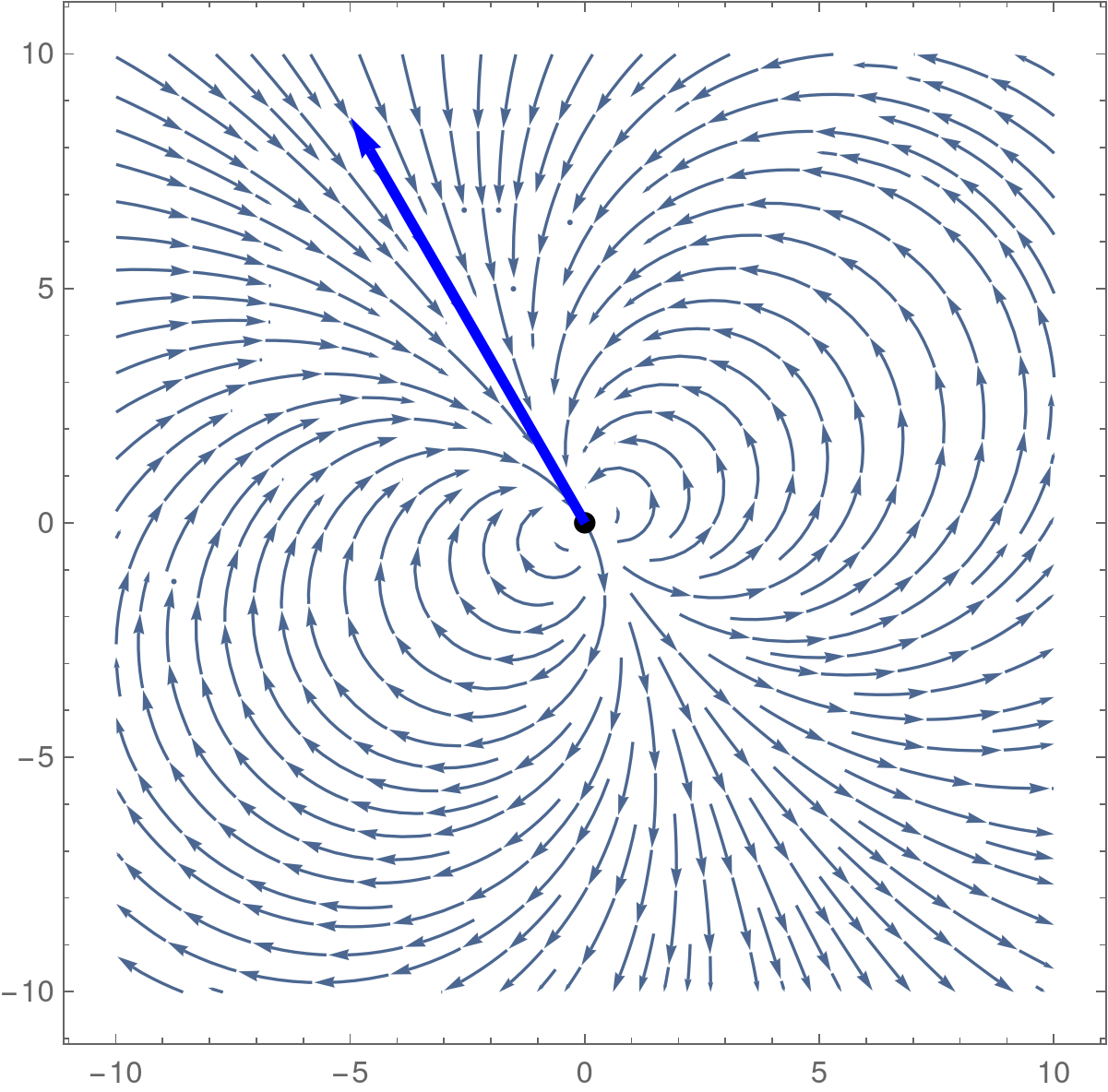}}
	\caption{Sketches of active forces with $\lambda>0$ for (a) two $+1$ defects, and (b), for a single $+2$ defect (the ``self-propulsion'' force). The forces for both cases are essentially in the same direction. 
	}
	\label{fig:Plus2}
\end{figure}

We now examine the net force. Since $I^{(1)}_{ij} \neq I^{(2)}_{ij}$, then $f_{ij}$ is a generic non-central force; in particular, it is also not orthogonal to the line connecting the two defects. We also comment that since $f_{ij} = f_{ji}$, then the defect pair moves together, as if it is a bound object. Another feature is that for a pair of $-1$ defects, there is no dependence on the distance between the defects, unlike in cases of the neutral pair or pair of $+1$ defects. See Fig.~\ref{fig:pair-forces} and Fig.~\ref{fig:Plus2} for sketches.

We have learned that two $+1$ defects exert the same force on each other (same magnitude and direction), as if they're bound. In the limit that these defects are really close to each other, then there is no reason a priori to expect that they are actually bound, as our assumptions no longer hold. However, interestingly enough, the two defects behave as if they're a $+2$ defect, a bound state of two $+1$ defects, which is  ``self-propelled" in the same direction, along its separatrix, consistent with the behavior of a $+2$ defect (see Fig.~\ref{fig:Plus2}). This did not have to be the case, and does not hold for the other defect pairs.

\section{Orientation dynamics}
\label{sec:pol}

\begin{figure}
	\includegraphics[scale=.5]{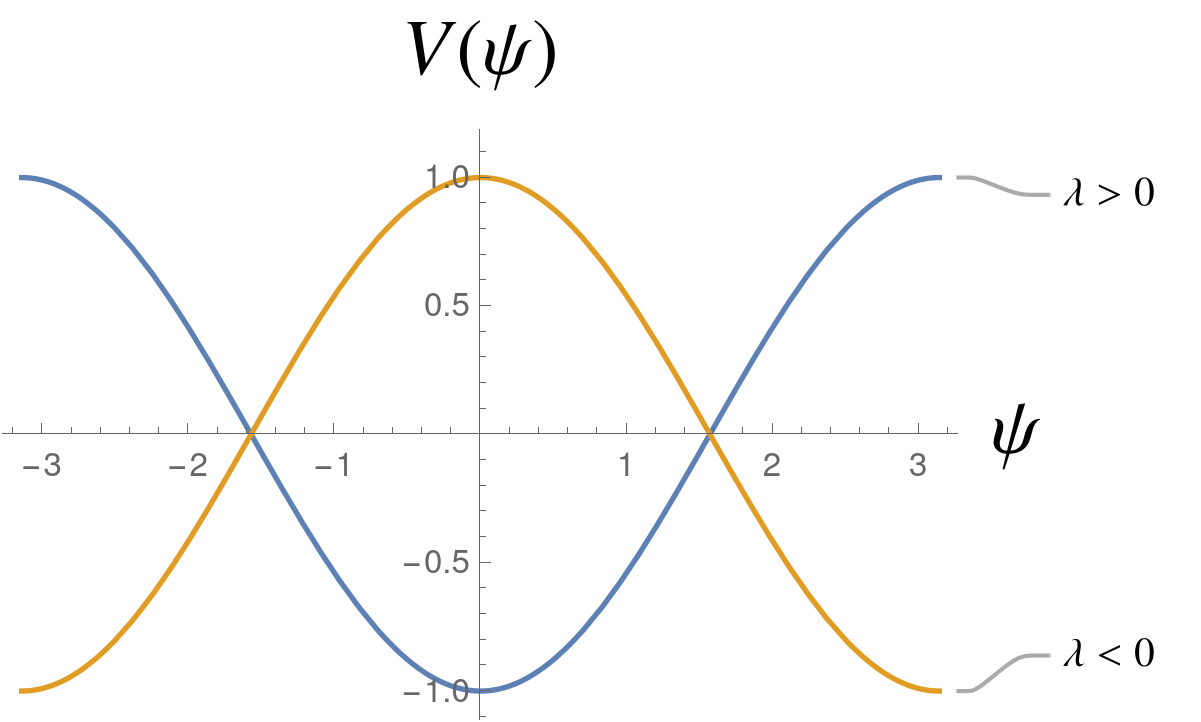}
	\caption{Plot of $V(\psi)$ for $\lambda>0$ and $\lambda<0$. Extrema are at $\psi=0,\pi$. Minimum for $\lambda>0$ is at $\psi=0$, whereas minimum for $\lambda<0$ is at $\psi=\pi$.}
	\label{fig:cospsi}
\end{figure}

In the previous section, we ignored orientation dynamics. We now incorporate orientation dynamics and sketch out the argument here (the details of the computation are in Appendix~\ref{app:pol}). For simplicity, we consider a single defect of charge $\sigma$ at the origin, in which case our ansatz is
\beq p_0 = e^{i\psi(t)} \left(\frac{z}{|z|}\right)^{\sigma}\; ,\eeq
where now the phase $\psi(t)$ is dynamical. Choosing $w_a(t) = \psi(t)$ in Eq.~\ref{eq:E} leads to
\beq \int d^2z |\p{p}{\psi}|^2 \dot \psi = \frac{\lambda}{2} \int d^2z \p{\bar p}{\psi} \mathcal I_\lambda + c.c \eeq
and upon evaluation in a region of size $\ell$ near the defect, where $a \ll \ell \ll L$ and $a$ is the core size,
\beq \pi \ell^2 \dot\psi = -2\pi\lambda \ell  \sin\psi \delta_{\sigma,1} \implies \dot\psi = -2\frac{\lambda}{\ell} \sin\psi \delta_{\sigma,1} \label{eq:psidot} \; .\eeq
Only the solution for $+1$ defects is nontrivial, which for completeness is given by
\beq \psi(t) = 2\,\rm{arccot}\left(e^{\frac{2\lambda}{\ell} t}\cot\left(\frac{\psi(0)}{2}\right)\right) \; . \eeq
Note that we can interpret Eq.~\ref{eq:psidot} as relaxational dynamics
\beq \dot\psi = -\frac{2}{\ell}\frac{dV}{d\psi} \label{eq:dVdpsi}\eeq
for the potential $V = -\lambda\cos\psi$ (see Fig.~\ref{fig:cospsi} for a plot). Thus for $\lambda>0$, the defect will relax to an aster ($\psi=0$), and for $\lambda<0$, the defect will relax to an inward-pointing aster ($\psi=\pi$).\footnote{Note that there is a symmetry of our system when $\lambda\to-\lambda$ and $p\to-p$ symmetry.}. In other words, there is a preferred phase. Stable asters have been observed in related simulations~\cite{Aranson2005pattern,Aranson2006theory,elgeti2011defect,Gopinath2012dynamical,Gowrishankar2016nonequilibrium,Husain2017emergent}, as well as analyzed in related models~\cite{Youn2001macroscopic,Sankararaman2004self,Kruse2004asters,Kruse2005generic,elgeti2011defect}.

\begin{figure}[]
	\centering
	\subcaptionbox{$+1$ defect}
	{\includegraphics[width=0.49\columnwidth]{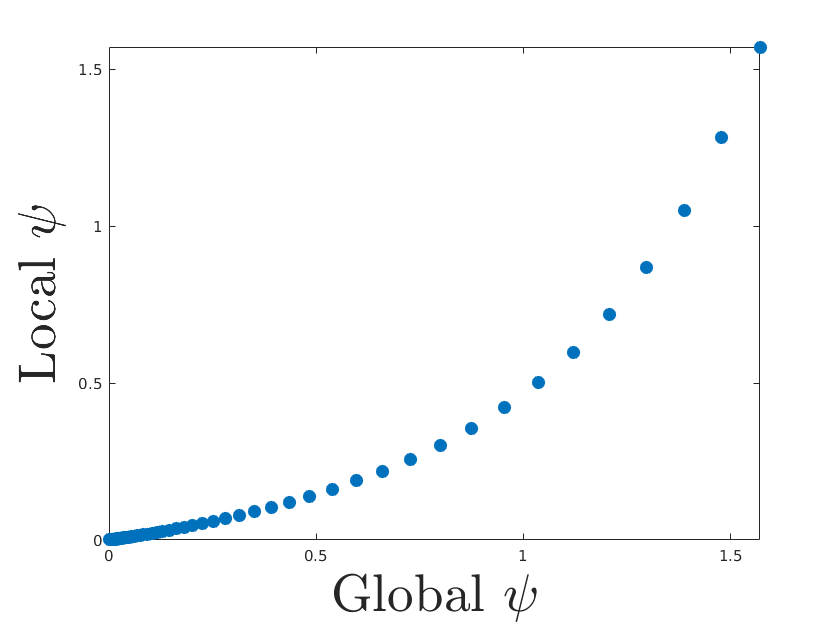}}
	\subcaptionbox{$-1$ defect}
	{\includegraphics[width=0.49\columnwidth]{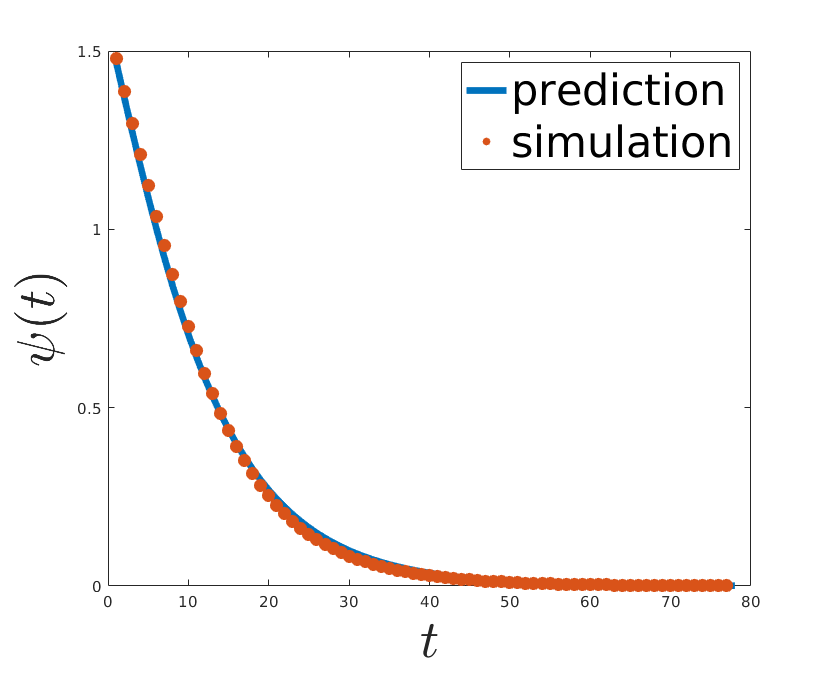}}
	\caption{Dynamics of the phase $\psi(t)$ of a single $+1$ defect for $\lambda>0$ with $\psi(0) = \pi/2$. In (a), plot of local computation vs global computation of the phase $\psi$ for a single $+1$. Their ending points are the same, but they are not identical in the middle. In (b), theoretical prediction vs simulation of $\psi(t)$.}
	\label{fig:simulation}
\end{figure}

We also check our theory with simulations. We evolve an isolated $+1$ defect for nonzero $\lambda = 1$, where initially the phase $\psi(0) = \pi/2$. We computed the phase in two different ways: a local computation, which locates the $+1$ defect and measures the phase, and a global computation, which calculates the defect position and phase by minimizing in a region of size $\ell = 30 a \ll L = 300 a$ the deviation of our ansatz $p_0$ from the measured $p_0$, which is basically equivalent minimizing Eq.~\eqref{eq:E}, as we did in deriving Eq.~\eqref{eq:psidot}. We find that initially and at long times, the two different measurements of the phase agree, and even though they are not identical in the middle, they both are similar. Moreover, we checked our measured global definition of $\psi(t)$ vs that predicted from theory obtained by integrating Eq.~\eqref{eq:dVdpsi}, and find remarkable agreement (see Fig.~\ref{fig:simulation}).

Given that our method suggests that there appear to be two different stationary solutions for $+1$ defects (aster or inward-pointing aster, depending on the sign of $\lambda$), it raises the question whether these solutions are stationary solutions of Eq.~\eqref{eq:complexp}. By inspection, $+1$ defects, in particular asters or inward-pointing asters, are indeed stationary solutions of Eq.~\eqref{eq:complexp}.

Since the phase appears to be important, it is natural to ask if we can modify our ansatz in Eq.~\eqref{eq:ansatz} to take into account the phase, for example by taking $\Psi_i \to e^{i\phi_i f(|z - z_i|)} \Psi_i$, where for example as in \cite{tang2017orientation} $f(|z - z_i|) = e^{i\gamma_i \ln |z - z_i|}$ \footnote{We do not assume this form of $f$ as this modified ansatz leads to an infinite free energy addition.}. We leave this analysis to future work.

\section{Stationary solution through scaling argument}
\label{sec:scaling}

In this paper, we have focused on defects. Here we make contact with the discussion contained in~\cite{Chate2020dry}, and provide another perspective about why defects are transient in active polar fluids.

We make use of a scaling argument. By inspection, there is a scaling symmetry; that is, solutions obey\footnote{For notational convience, we drop the explicit dependence on $g$. Explicitly, $g$ scales as $p(z,t; g, \lambda) = p(z/\beta, t/\beta^2;\beta^2 g, \beta\lambda)$. Since we are in the deep nematic limit, $g\to\infty$, so it is unaffected by rescaling. But for finite $g$, this is how it would scale.}
\beq p(z,t; \lambda) = p(z/\beta, t/\beta^2; \beta\lambda)\label{eq:scaling} \; .\eeq
We are interested in the stationary, longtime behavior, which means that we are looking for $p$ such that for any $t$ 
\beq \lim_{\gamma\to\infty}\p{}{t} p(z,\gamma^2t;\lambda) = 0 \; .\eeq
From our scaling relation in Eq.~\ref{eq:scaling}, choosing $\beta=\gamma$ is equivalent to finding $p$ such that
\beq \lim_{\gamma\to\infty}\p{}{t} p(z/\gamma,t;\gamma\lambda) = 0 \; .\eeq
We thus look for steady states for large $\lambda$. For large $\lambda$, the advection term in Eq.~\eqref{eq:complexp} dominates, and thus long-time stationary states satisfy
\beq \partial_t p = -\lambda (p \partial + \bar p \bar\partial) p = 0 \; .\eeq
We will now show that the only solutions to the above equation other than constant $p$ is a single aster or inward-pointing aster, which as we commented in Sec.~\ref{sec:pol} satisfies the above equation. Because we are deep in the ordered phase, our ansatz is $p = \frac{f(z)}{\bar f(\bar z)}$. Then 
\beq \partial_tp = -\lambda(p\partial + \bar p \bar\partial)p = - \frac{\lambda}{\bar f^2}(f \partial f - \bar f\bar \partial \bar f)\eeq
which vanishes only if $\partial (f^2) = c_1$, where $c_1 \in \mathbb{R}$. Therefore, $f^2 = c_1z + c_2$, and so $p$ is constant if $c_1 = 0$, and otherwise
\beq p = e^{i\psi}\frac{(z - z_i)^{1/2}}{(\bar z - \bar z_i)^{1/2}}\label{eq:exactPlus1}\eeq
where either $\psi = 0$ (aster) or $\psi = \pi$ (inward pointing aster), depending on the sign of $\lambda$; no other $\psi$ is allowed. Note that this single aster stationary state is consistent with the single vortex to aster transition, as in Eq.~\eqref{eq:dVdpsi}. We have thus provided another perspective for transient behavior of defects.

\section{Comparison with active nematics model}
\label{sec:comparison}

\subsection{Overview}

In this section, we compare our model to the active nematics model studied in~\cite{vafa2020multidefect}. We first present a general overview, and then study the consequences. The similarities are that both models advect an order parameter deep in the ordered phase and in the overdamped limit. In the case of nematic, the order parameter is a rank 2 symmetric traceless tensor $\mathbf Q$, and in the case of polar, the order parameter is a vector. This difference implies that there are extra terms in the advection of $\mathbf Q$. In the case of nematic, overdamped limit implies $\mathbf v = \alpha \nabla\cdot\mathbf Q$, where $\alpha$ is a measure of activity, and in the case of polar, $\mathbf v = \lambda \mathbf p$. This difference in dependence of length scaling implies that in the nematic model, $\alpha$ cannot be scaled out of the problem, but in the polar model, $\lambda$ can be scaled out. Although these models are different, they are similar, and by studying these models in depth it is interesting to learn which features are common and which are model-dependent.

\subsection{Forces}

We now compare the forces. In the absence of activity, the models are equivalent. The forces that arise because of activity are different. In the active nematics case, a $+1/2$ defect, the smallest allowed energy excitation, is ``self-propelled", whereas in the active polar case, a $\pm 1$ defect, the smallest allowed energy excitation, is not ``self-propelled"; a $+2$ defect is ``self-propelled". Another difference between these two models arise in the pair-wise interactions induced by activity. In the active nematics case, the active forces are central for a $(+1/2,+1/2)$ pair, and for the other pairs are orthogonal to line connecting the defects. Also, the forces for $(+1/2,-1/2)$ pair are non-reciprocal. All of these forces fall off as $1/r$, where $r$ is the distance between the defects, and the magnitude depends on the geometry, that is, overall phase of $Q$. In contrast, in the case of active polar, the active forces are neither central forces nor orthogonal to the line connecting the two defects. They are also always equal, and except for the $-1$ defect pair, goes as $\ln L/r$. Similar to active nematic, the magnitude of the force depends on the geometry, that is, the phase of $p$.

\subsection{Orientation dynamics / solutions}

\begin{figure*}[]
	\centering
	\subcaptionbox{Aster bound state}
	{\includegraphics[width=0.32\textwidth]{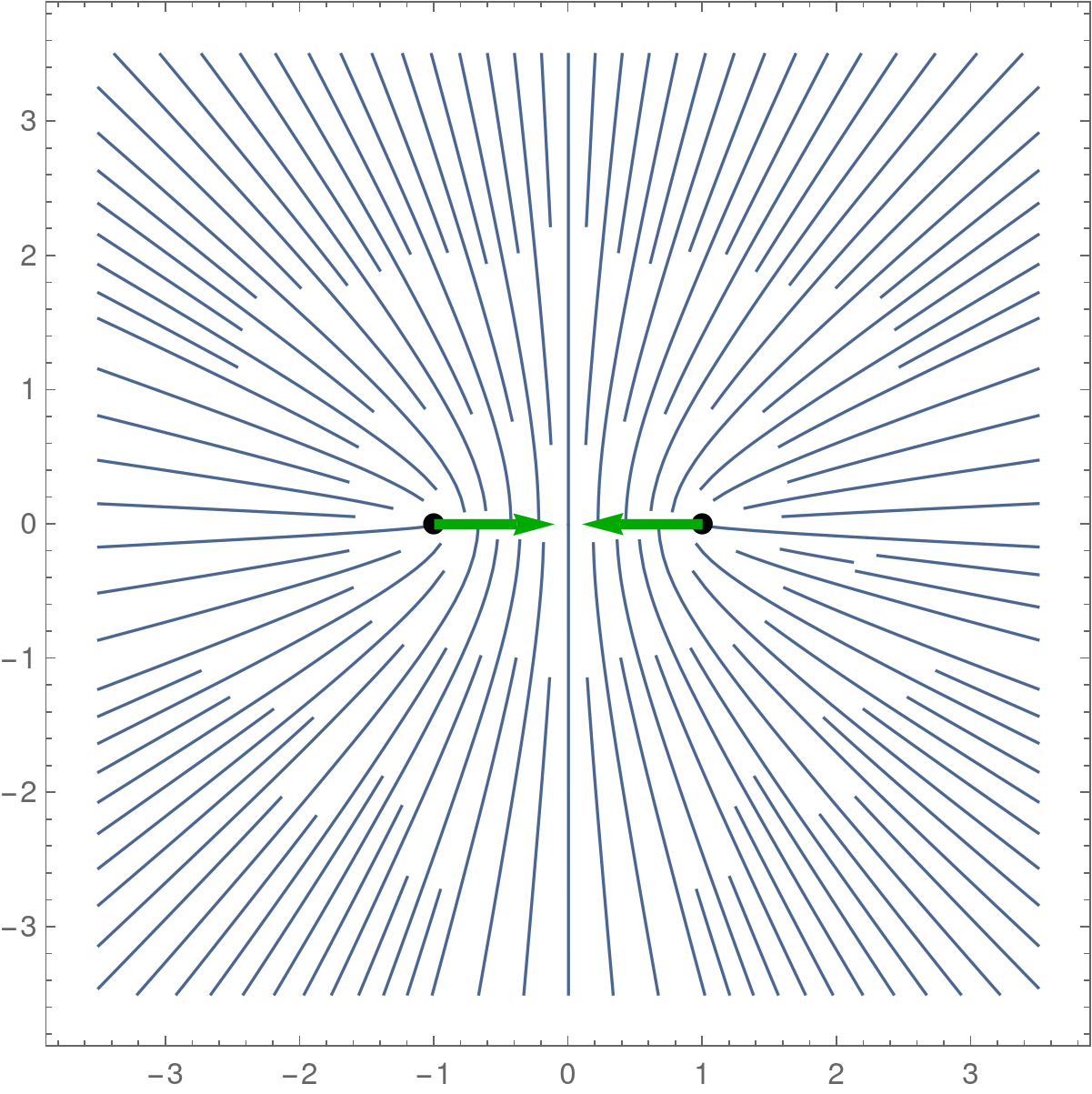}}
	\subcaptionbox{Vortex bound state}
	{\includegraphics[width=0.32\textwidth]{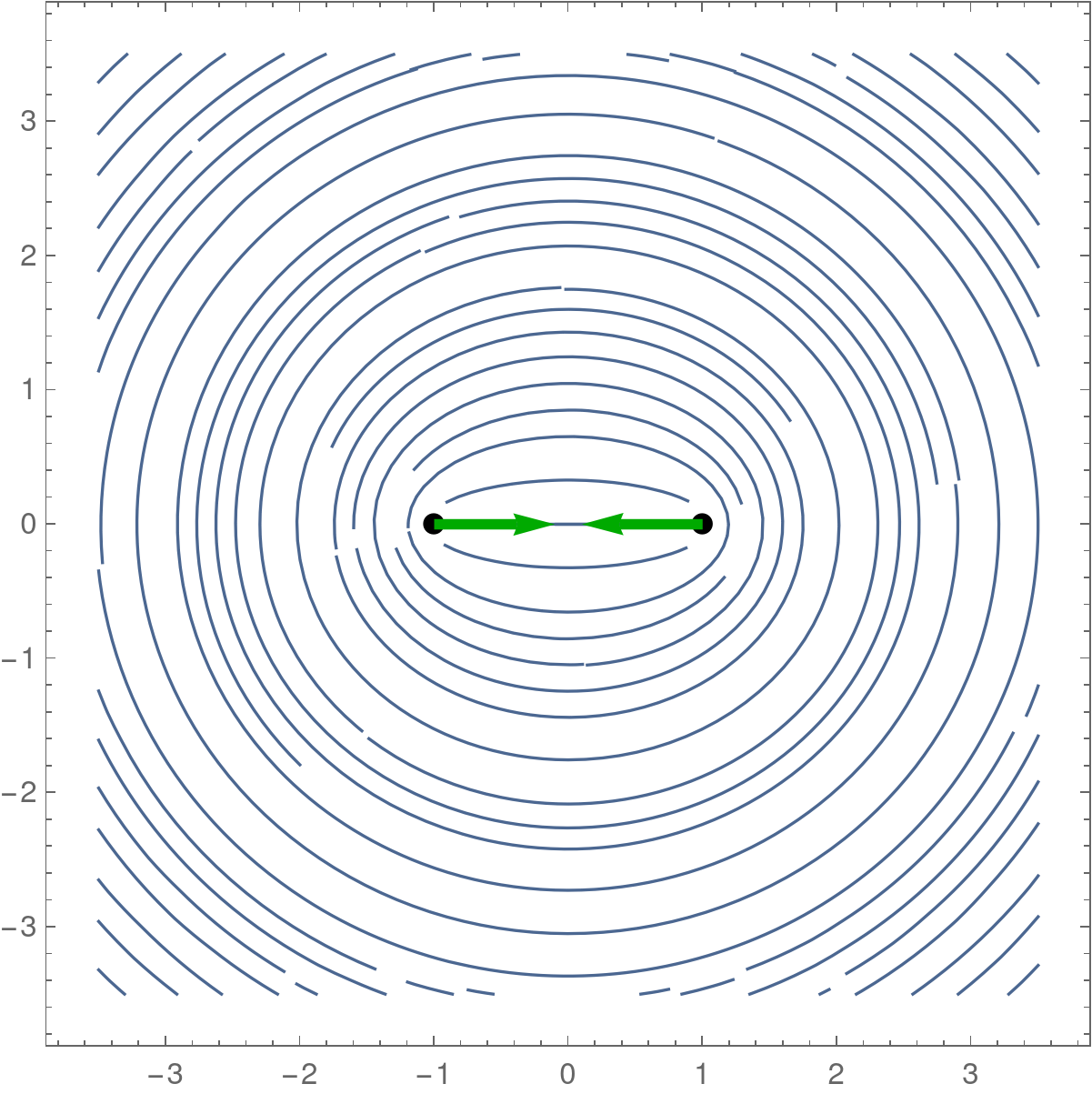}}
	\subcaptionbox{Spiral bound state}
	{\includegraphics[width=0.32\textwidth]{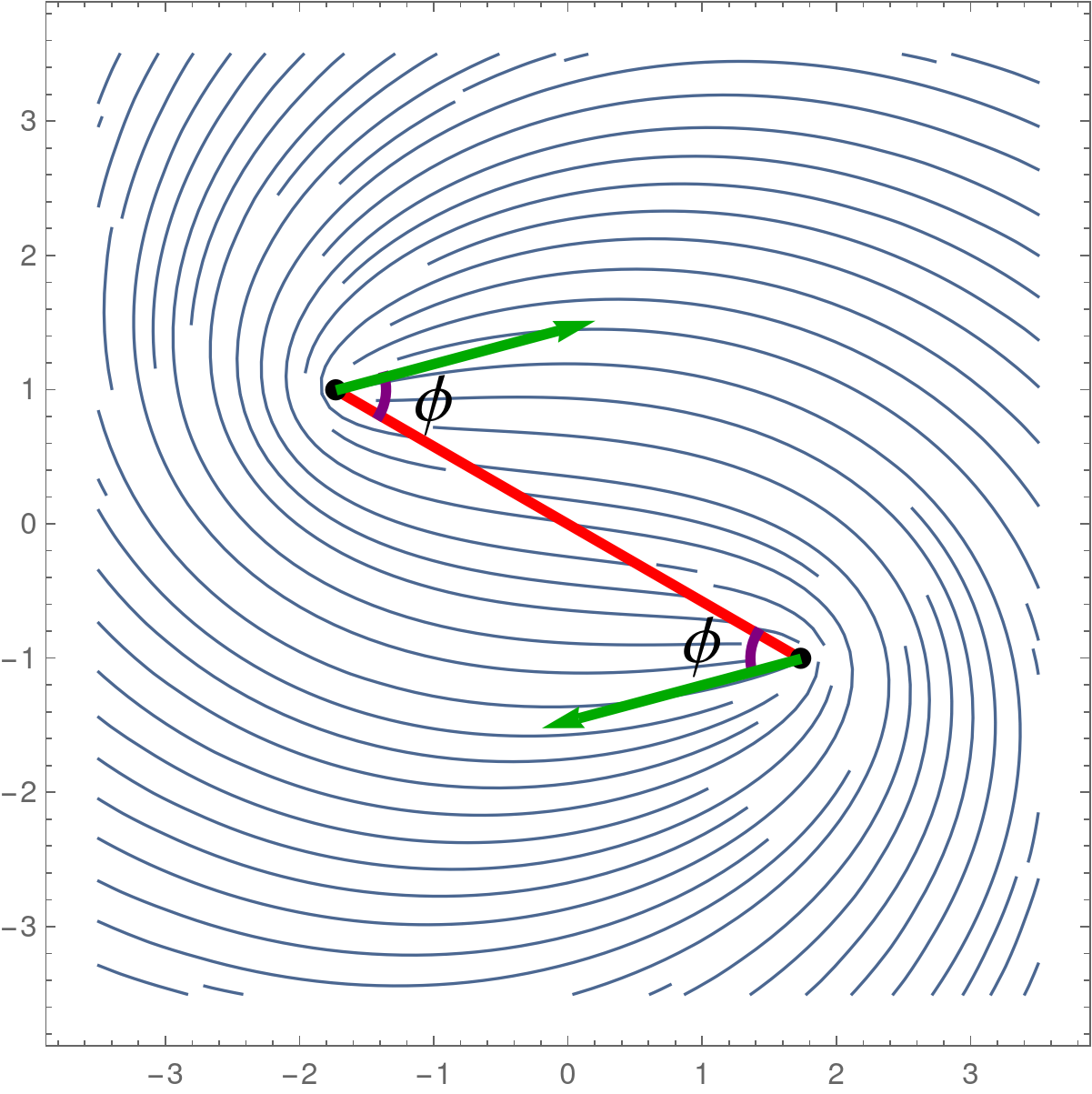}}
	\caption{In active nematics model, sketches of bound state of two $+1/2$ defects. In (a), bound aster state in extensile system when $\psi=0$, (b), bound vortex state in contractile system when $\psi = \pi$, and in (c), bound spiral state in contractile system when $0<\psi<\pi/2$.}
	\label{fig:boundStates}
\end{figure*}

In this paper, we learned that $+1$ asters (inward-pointing asters) are stationary solutions and that they are stable for $\lambda>0$ ($\lambda < 0$). It is natural to ask whether in the nematic model there can be stationary $+1$ defect configurations, and does the existence of solutions, or stability, depend on the phase of the defects. We show that indeed solutions exist, and the type of solution depends on the phase of the defects. 

We first check to see what happens if we incorporate orientation dynamics into the active nematics model. The active nematics model has the following equation of motion,
\beq
\partial_t Q = \mathcal I(Q) = -\frac{\delta {\mathcal F}(\{Q\})}{\delta {\bar Q}}+\alpha \mathcal I_\alpha(Q)\;,
\label{eq:complexQ}
\eeq
where
\begin{gather}
\frac{\delta {\mathcal F}(\{Q\})}{\delta {\bar Q}} = -4 {\bar \partial }\partial Q - 2\epsilon^{-2}(1- |Q|^2)Q\\
\mathcal I_\alpha (Q) = -(\partial Q\partial Q + {\bar \partial} {\bar Q}{\bar \partial} Q) + (\partial^2 Q- \bar\partial^2\bar Q) Q
\end{gather}
We work in the deep nematic limit ($\epsilon\to 0$). For simplicity, we consider a single defect of charge $\sigma=\pm 1/2$ at the origin, in which case our ansatz is
\beq Q_0 = e^{i\psi(t)} \left(\frac{z}{\bar z}\right)^{\sigma}\; ,\eeq
where now the phase $\psi(t)$ is dynamical. Minimizing the error
\begin{align}
E &= \int d^2dz d\bar{z} \left| \partial_t Q( z,{\bar z},t) - \frac{d}{dt} Q_0( z,{\bar z}|\psi(t))\right|^2 \nonumber\\
&\approx \int dz d\bar{z} \left| \mathcal I (Q_0) - \dot \psi\p{Q_0}{\psi} \right|^2
\end{align}
with respect to $\dot\psi$ (the analogue of Eq.~\eqref{eq:E}) leads to
\beq \int d^2z |\p{Q_0}{\psi}|^2 \dot \psi = \frac{\alpha}{2} \int d^2z \p{\bar Q_0}{\psi} \mathcal I_\alpha + c.c \eeq
(the analogue of Eq.~\eqref{eq:ODE}).  We now evaluate both sides of this equation in a region near the defect of size $\ell$. As before,
\beq \int d^2z |\p{Q_0}{\psi}|^2 = \pi \ell^2\eeq
We now evaluate the RHS. We have
\begin{align}  & \int d^2z \p{\bar Q_0}{\psi} \mathcal I_\alpha + c.c = \\
&-i\int d^2z[-\sigma^2(\frac{Q_0}{z^2} - \frac{\bar Q_0}{\bar z^2} ) + \sigma(\sigma-1)(\frac{Q_0}{z^2} - \frac{\bar Q_0}{\bar z^2})] + c.c \nonumber\\
&= 0
\end{align}
which implies that
\beq \dot \psi = 0\eeq
and we thus learn that the phase is frozen, in accordance with the expectation in~\cite{vafa2020multidefect}. Here there is no preferred orientation, unlike in the active polar case, where asters or anti-asters are preferred, depending on the sign of $\lambda$.

In related models, $+1$ defect states consisting of two $+1/2$ defects have been observed in active nematics \cite{kumar2018tunable,Pearce2020,Thijssen2020}, and in~\cite{shankar2018defect,Thijssen2020binding}, it was argued that the type of $+1$ defect was determined by the activity: asters in extensile systems, and vortices in contractile systems. This observation is related to our result of finding a stationary $+1$ defect in the active polar model, as we will now see. We now review and present another argument for the existence and stability of a stationary defect pair of two $+1$ defects in the active nematics case.

Let's consider two $+1/2$ defects situated on the real axis. The orientations of the $+1/2$ defects anti-align~\cite{vromans2016orientational,Pearce2020,Thijssen2020,vafa2020multidefect}. For simplicity, let's assume that the orientations are along the real axis, so they either point away from each other (phase is 0), or toward each other (phase is $\pi$). There are four forces: the defect drag force, the repulsive Coulomb force, the self-propulsion, and the active induced pair-wise force. We will ignore the defect drag force and active induced pair-wise force because they renormalize the velocity and Coulomb force, respectively. In this case, for $\alpha>0$ (contractile), the $+1/2$ defects move with constant velocity in the direction of their phase, and for $\alpha<0$ (extensile), the $+1/2$ defects move with constant velocity in the opposite direction of their phase. Therefore, at a unique separation $r_*$ the repulsive Coulomb force can balance the attractive self-propulsion force depending on the sign of $\alpha$ and the phase. The configuration is stationary for either extensile system and phase is $0$ or contractile system and phase is $\pi$. In the former, the two $+1/2$ defects form a bound aster state, and in the latter, they form a bound vortex state (see Fig.~\ref{fig:boundStates}). This argument was pointed out in~\cite{shankar2018defect,Turiv2020topology,Thijssen2020binding}.

Moreover, this bound state is stable to transverse fluctuations of the polarization~\cite{shankar2018defect}. Here we present an alternative argument. If the defects are not exactly aligned, one would naively think that the self-propulsion will cause the $+1/2$ defects to go away from each other. However, we will now show that as the defects move, the orientation readjusts in such a way that it leads to inward spiral motion of the pair of defects. From arguments presented in~\cite{vafa2020multidefect}, in terms of this phase $\phi$ (the angle of the orientation, that is, the deviation from radial line connecting the two defects), the solution takes the form
\beq Q(z,t) = e^{i\phi} \frac{z - z_i(t)}{|z - z_i(t)|}\frac{z - z_j(t)}{|z - z_j(t)|}\eeq
where $z_i$ and $z_j$ are the positions of defects $i$ and $j$, respectively. The orientation $Q_i(t)$ of defect $i$ is simply
\beq Q_i(t) = e^{i\phi}\frac{z_i(t) - z_j(t)}{|z_i(t) - z_j(t)|}\eeq
Since $+1/2$ defects are self-propelled along their orientation, in the direction of $Q_i$, then they will always move at a constant angle $\phi$ relative to the radial line connecting the two defects. Thus for example in contractile system, if $\phi$ is sufficiently close to $0$, and the activity is not too large, then the two defects will simply spiral towards each other (see Fig.~\ref{fig:boundStates}). The solution is thus stable, but not stationary.

\begin{figure}[]
	\centering
	\subcaptionbox{Asters}
	{\includegraphics[width=.45\columnwidth]{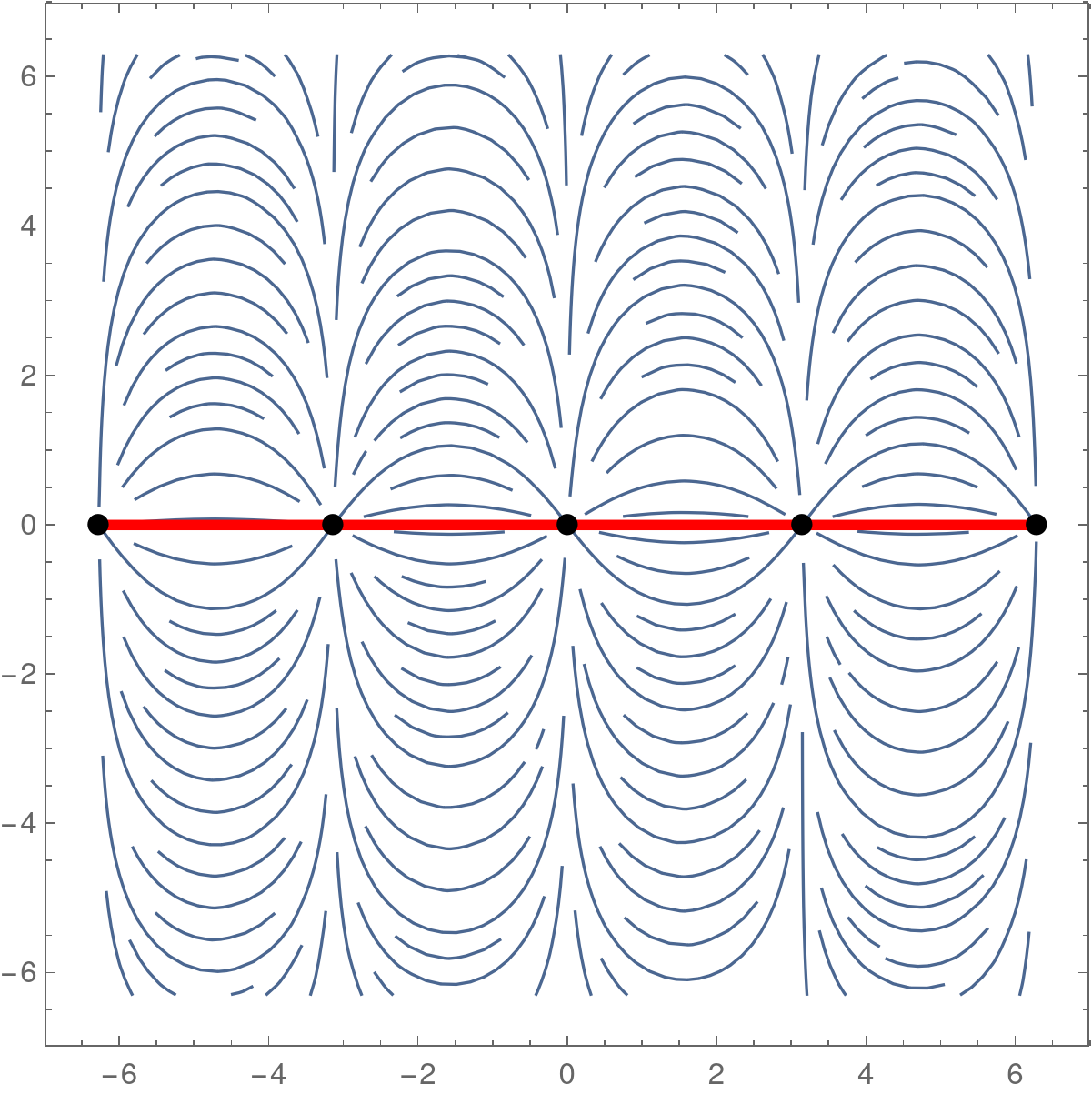}}
	\subcaptionbox{Vortices}
	{\includegraphics[width=.45\columnwidth]{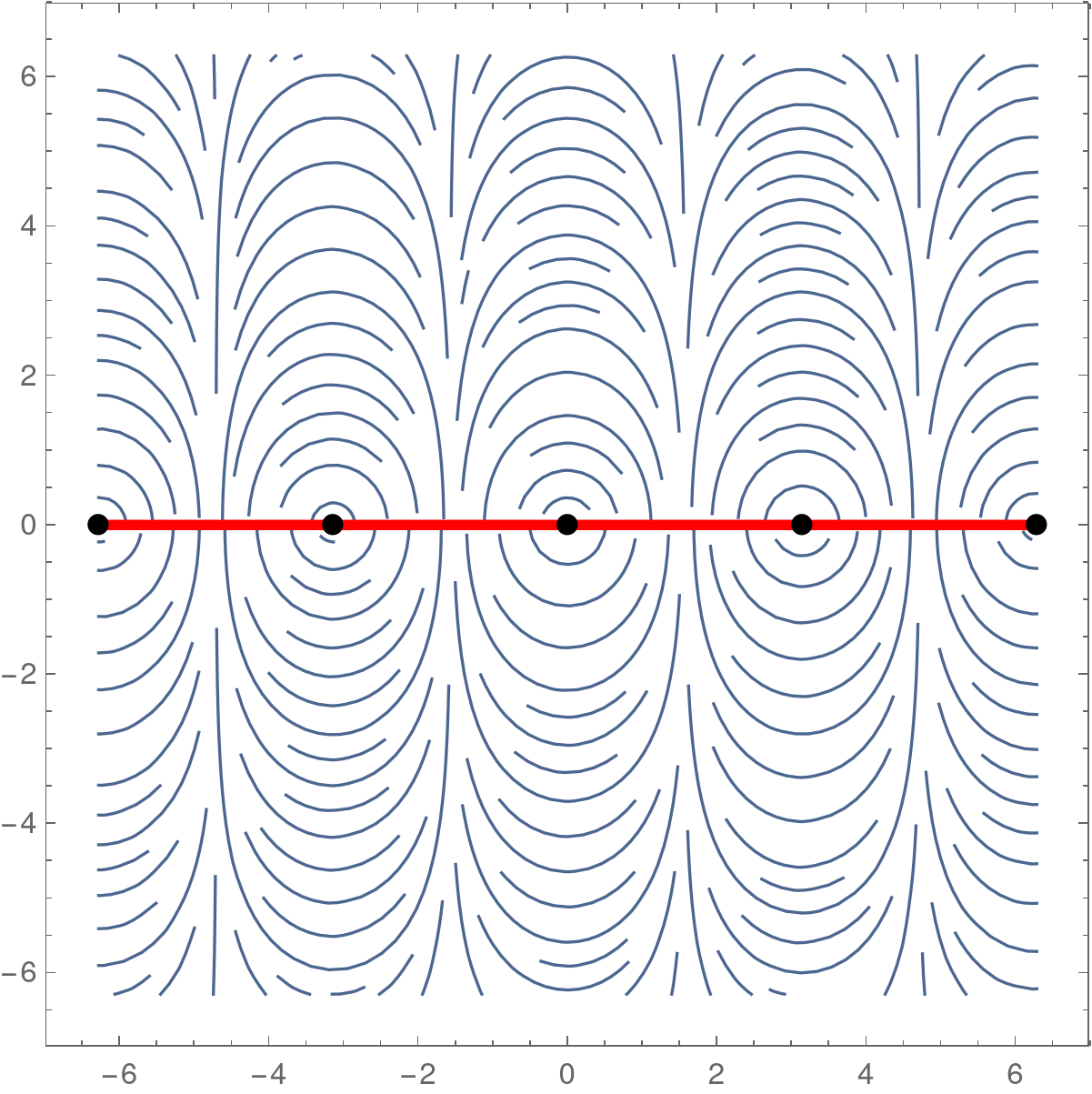}}
	\caption{Configuration of 1D chain of equally spaced $+1$ defects for active nematics model that screens the activity term.
	}
	\label{fig:chain}
\end{figure}

Given that it seems that a composite made of a pair of $+1/2$ defects is a stationary solution for the active nematic model and far away it looks like an aster or vortex, it is natural to ask if an aster or vortex is actually a solution to Eq.~\eqref{eq:complexQ}. By inspection, indeed a nontrivial solution is $Q = \pm\frac{z}{\bar z}$ (as one can easily check that the active term $\mathcal I_\alpha=0$), where the $+$ sign corresponds to an aster and the $-$ sign corresponds to a vortex.
Note that this solution of aster or vortex is consistent with the picture in Fig.~\ref{fig:boundStates}, as any other phase results in a non-stationary state. Thus a single aster or a vortex is indeed a stationary solution to Eq.~\eqref{eq:complexQ}.

Screening of activity term by $+1$ defects in active nematics is similar to what we found in active polar fluids. In active polar fluids, this is the only configuration which screens the active term and that is the reason for transient behavior of defects. Is this the case in active nematics or are there more general configurations that screen the active term? Or can we extend this solution to allow multiple defects? A natural place to look for this (ignoring the passive forces) is to look for configurations that screen the active term ($\mathcal I_\alpha=0$), as in the case of single aster/vortex. In the polar case, a single aster was the only defect configuration that screened the active term. Here we will see that the situation (and solution) is more interesting for a nematic system.

We are thus interested in solving
\beq \mathcal I_\alpha = 0 \; ,\eeq
where
\beq \mathcal I_\alpha =  -\partial Q\partial Q - {\bar \partial} {\bar Q}{\bar \partial} Q + (\partial^2 Q- \bar\partial^2\bar Q) Q\eeq

Deep in the ordered phase, $Q = \pm e^{i(f(z) + \bar f(\bar z))}$, and so
\beq iQ\left(\partial^2f e^{2if} +\bar\partial^2\bar f e^{-2i\bar f}\right) = 0\eeq
Other than the constant solution, the unique solution is
\beq f(z) = -i\ln\sin k(z - z_0)\eeq
where without loss of generality we can assume $k\in\mathbb{R}$ by rotation of $z$ coordinate if necessary and place the origin at $z_0$. Therefore,
\beq Q = \pm e^{i(f(z) + \bar f(\bar z))} = \pm \frac{\sin k z}{\sin k \bar z} \; .\eeq

Notice that this vanishes at $z = n \pi/k$, for $n\in \mathbb Z$, and near each zero, $Q\sim \pm \frac{z}{\bar z}$. We thus have an infinite chain of $+1$ nematic defects on the real axis, separated by $\pi/k$. Because of the sign of $Q$, either the defects are all asters (when the sign is positive), or the defects are all vortices (when the sign is negative). These configurations are depicted in Fig.~\ref{fig:chain}.

Ignoring the Coulomb term, we have analytically found a stationary $1D$ lattice solution. For example, in the geometry of a thin annulus (or equivalently, long channel with periodic boundary conditions), we can imagine that the boundary condition balances the Coulomb forces. 
In any case, this shows that $\mathcal I_\alpha = 0$ has a much more interesting set of solutions than $\mathcal I_\lambda = 0$, and deserves further study, pointing to the importance of defects in active nematic systems as opposed to active polar systems.

\begin{acknowledgments}
	
	I would like to thank M.~Cristina Marchetti for many valuable comments on this manuscript. In addition, I have benefited from discussions with Mark Bowick, Sattvic Ray, and Boris Shraiman.
	
	This work was supported in part by the NSF through grants DMR-1938187 and PHY-0844989.
	
\end{acknowledgments}

\section*{Appendices}

\appendix

\section{Single defect solution}
\label{app:A}

Stationary textures in the limit of zero activity ($\lambda =0$) minimize  free energy and hence solve ~\cite{de1993physics,Pismen1999} 
\beq
\frac{\delta {\mathcal F}}{\delta {\bar p}}= -\nabla^2 p - 2\epsilon^{-2}(1- |Q|^2)p =0\;.
\eeq
We look for a solution for a single defect of charge $\sigma$ of the form
\beq p = A(r)e^{i\sigma\varphi}\;.\eeq
$A(r)$ would thus satisfy
\beq A''(r) + \frac{A'}{r} + \left(2\epsilon^{-2} - \frac{\sigma^2}{r^2} - 2\epsilon^{-2} A^2\right)A = 0\;.\label{A}\eeq
For example, for $\sigma = \pm 1$, $A(r)$ can be approximated as \cite{Pismen1999}
\beq A(r) = \tilde r \sqrt{\frac{.68 + .28\tilde r^2}{1 + .82 \tilde r^2 + .28 \tilde r^4}}\label{ASoln}\;,\eeq
where $\tilde r = r/\epsilon$. As $r\to0$, $A(r) \propto r$, and for $r \gg \epsilon$, $A(r) \simeq 1 - \frac{\epsilon^2}{4r^2}$. The defect core size $a$, which is the length scale over which $A$ goes from 0 to 1, is of the order $a \sim \epsilon$.

\section{Computation of $\mathcal U_i$}
\label{app:U_i}

We are interested in computing
\beq \mathcal U_i = \int d^2z \bar \partial_i \bar p_0 \mathcal I_\lambda + \int d^2z \bar \partial_i p_0 \bar{\mathcal I}_\lambda = I_1 + I_2\;,\eeq
where
\begin{align}
I_1 &= -\int d^2z \bar \partial_i \bar p_0 p_0 \partial p_0 - \int d^2z \bar \partial_i p_0 p_0 \partial \bar p_0\\
I_2 &= -\int d^2z \bar \partial_i p_0 \bar p_0 \bar \partial \bar p_0 - \int d^2z \bar \partial_i \bar p_0 \bar p_0 \bar \partial p_0 \;.
\end{align}
Substituting for $p_0$, we find that
\begin{align}
I_1 &= \frac{1}{2} \sum_j \int d^2z p_0 \frac{\sigma_i\sigma_j}{(\bar z - \bar z_i)(z - z_j)}\\
I_2 &= -\frac{1}{2} \sum_j \int d^2z \bar p_0 \frac{\sigma_i\sigma_j}{(\bar z - \bar z_i)(\bar z - \bar z_j)}
\end{align}
It is convenient to rewrite the above as
\begin{align}
I_1 &= \frac{1}{2}\int d^2z p_0 \frac{\sigma_i^2}{|z - z_i|^2} + \frac{1}{2} \sum_{j\neq i} \int d^2z p_0 \frac{\sigma_i\sigma_j}{(\bar z - \bar z_i)(z - z_j)}\\
I_2 &= -\frac{1}{2} \int d^2z \bar p_0 \frac{\sigma_i^2}{(\bar z - \bar z_i)^2} -\frac{1}{2} \sum_{j\neq i} \int d^2z \bar p_0 \frac{\sigma_i\sigma_j}{(\bar z - \bar z_i)(\bar z - \bar z_j)}
\end{align}
The first term for $I_1$ vanishes by phase integral. The first term for $I_2$ is only non-zero for $\sigma_i = 2$, in which case (since we are assuming that defects are well-separated), we can approximate
\begin{align}
-\frac{1}{2} \int d^2z \bar p_0 \frac{\sigma_i^2}{(\bar z - \bar z_i)^2}  &\approx -\delta_{\sigma_i,2}\frac{1}{2} \bar P_i \int d^2z \frac{(\bar z - \bar z_i)^2}{|z - z_i|^2} \frac{4}{(\bar z - \bar z_i)^2}  \nonumber \\
&= - 8 \pi \bar P_i \delta_{\sigma_i,2}\ln \frac{L}{a}
\end{align}
which we can identify as the self-propulsion of a $+2$ defect. In the following, we will explicitly be assuming that $\sigma_i = \pm 1$, so this term does not appear. Thus we can write
\begin{align}
I_1 &\approx \frac{1}{2}\sum_jP_{ij} \int d^2z \frac{(z - z_i)^{\sigma_i}}{|z - z_i|^{\sigma_i}} \frac{(z - z_j)^{\sigma_j}}{|z - z_j|^{\sigma_j}} \frac{\sigma_i\sigma_j}{(\bar z - \bar z_i)(z - z_j)}\\
I_2 &\approx -\frac{1}{2} \sum_j \bar P_{ij} \int d^2z \frac{(\bar z - \bar z_i)^{\sigma_i}}{|z - z_i|^{\sigma_i}} \frac{(\bar z - \bar z_j)^{\sigma_j}}{|z - z_j|^{\sigma_j}} \frac{\sigma_i\sigma_j}{(\bar z - \bar z_i)(\bar z - \bar z_j)} \; ,
\end{align}
where
\beq P_{ij} =  \prod_{r\neq i,j}\frac{(z_i - z_r)^{\sigma_r}}{|z_i - z_r|^{\sigma_r}}\;.\eeq
First shifting $z \to z + z_j$ and then rescaling  $z\to z_{ij} z$, we have
\begin{align}
I_1 &= \frac{1}{2} \sum_j \sigma_i\sigma_j P_{ij} \left(\frac{z_{ij}}{|z_{ij}|}\right)^{\sigma_i + \sigma_j} I^{(1)}_{ij}\\
I_2 &= -\frac{1}{2} \sum_j \sigma_i\sigma_j \bar P_{ij} \left(\frac{\bar z_{ij}}{|z_{ij}|}\right)^{\sigma_i + \sigma_j-2} I^{(2)}_{ij}
\end{align}
where
\begin{align}
I^{(1)}_{ij} &= \int d^2z \frac{(z - 1)^{\sigma_i}}{|z - 1|^{\sigma_i}}\frac{z^{\sigma_j}}{|z|^{\sigma_j}}\frac{1}{\bar z - 1}\frac{1}{z}\\
I^{(2)}_{ij} &= \int d^2z \frac{(\bar z - 1)^{\sigma_i}}{|z - 1|^{\sigma_i}}\frac{\bar z^{\sigma_j}}{|z|^{\sigma_j}}\frac{1}{\bar z - 1}\frac{1}{\bar z}
\end{align}
are integrals that need to be computed. For notation, let $+(-)$ index denote plus (minus) defect. Using techniques utilized in~\cite{vafa2020multidefect}, we find that
\begin{gather}
I^{(1)}_{++} = I^{(1)}_{--} = 2\pi \nonumber\\
I^{(1)}_{+-} = I^{(1)}_{-+} =  2\pi \ln \frac{L}{r_{ij}} + \mathcal O(L^0) \nonumber\\
I^{(2)}_{++} = 2\pi \ln \frac{L}{r_{ij}} + \mathcal O(L^0); \quad I^{(2)}_{--} = 0 \nonumber\\
I^{(2)}_{+-} = I^{(2)}_{-+} = 2\pi
\end{gather}

To summarize, $\mathcal U_i$ can be written explicitly in terms of the defect positions as
\beq \lambda \mathcal U_i = -8\pi\ln \frac{L}{a}\lambda \bar P_i \delta_{\sigma_i,2} + \sum_{j\neq i} f_{ij},\eeq
where
\beq f_{ij} = \frac{1}{2}\sigma_i\sigma_j \left(P_{ij} \hat z_{ij}^{\sigma_i + \sigma_j}I^{(1)}_{ij} - \bar P_{ij} \hat {\bar z}_{ij}^{\sigma_i + \sigma_j-2}I^{(2)}_{ij}\right)\eeq
can be interpreted as the active induced pair-wise force on defect $i$ due to defect $j$. $f_{ij}$ can be rewritten as
\beq f_{ij} = \frac{1}{2}\sigma_i\sigma_j \hat z_{ij} \left(P_{ij} \hat z_{ij}^{\sigma_i + \sigma_j-1}I^{(1)}_{ij} - \bar P_{ij} \hat {\bar z}_{ij}^{\sigma_i + \sigma_j-1}I^{(2)}_{ij}\right)\eeq
or equivalently as
\beq f_{ij} = \frac{1}{2}\sigma_i\sigma_j \hat z_{ij} \left(P_i \hat z_{ij}^{\sigma_i -1}I^{(1)}_{ij} - \bar P_i \hat {\bar z}_{ij}^{\sigma_i -1}I^{(2)}_{ij}\right)\eeq

\section{Orientation dynamics computations}
\label{app:pol}
 For simplicity, we consider a single defect of charge $\sigma$ at the origin, in which case our ansatz is
\beq p_0 = e^{i\psi(t)} \left(\frac{z}{|z|}\right)^{\sigma}\; ,\eeq
where now the phase $\psi(t)$ is dynamical. Choosing $w_a(t) = \psi(t)$ in Eq.~\ref{eq:E} leads to
\beq \int d^2z |\p{p_0}{\psi}|^2 \dot \psi = \frac{\lambda}{2} \int d^2z \p{\bar p_0}{\psi} \mathcal I_\lambda(p_0) + c.c, \eeq
where the Coulomb term vanishes because there is only one defect. We now evaluation both sides of the above equation in a region of size $\ell$ near the defect, where $a \ll \ell \ll L$ and $a$ is the core size. We first evaluate the LHS. Since $|\p{p_0}{\psi}| = 1$, then
\beq \int d^2z |\p{p_0}{\psi}|^2 = \pi \ell^2 \; .\eeq
We now evaluate the RHS. We have
\begin{align}
\frac{\lambda}{2} \int d^2z \p{\bar p_0}{\psi} \mathcal I_\lambda(p_0) + c.c &= -\frac{\lambda}{2} \int d^2z \p{\bar p_0}{\psi} (p_0 \partial + \bar p_0 \bar \partial)p_0 + c.c \nonumber\\
&=  i\frac{\lambda}{2} \frac{\sigma}{2}\int d^2z (\frac{p_0}{z} - \frac{\bar p_0}{\bar z}) + c.c \; .
\end{align}
By phase integral, the above vanishes unless $p_0 = e^{i\psi} \frac{z}{|z|}$, that is, $\sigma = 1$. Thus
\begin{align}
\frac{\lambda}{2} \int d^2z \p{\bar p_0}{\psi} \mathcal I_\lambda(p_0) + c.c &=  -\frac{\lambda}{2} \sin\psi \delta_{\sigma,1}\int d^2z \frac{1}{|z|} + c.c \nonumber\\
&= -2\pi\lambda \ell \sin\psi \delta_{\sigma,1} \; .
\end{align}
Putting it all together,
\beq \pi \ell^2 \dot\psi = -2\pi\lambda \ell  \sin\psi \delta_{\sigma,1} \implies \dot\psi = -2\frac{\lambda}{\ell} \sin\psi \delta_{\sigma,1} \; .\eeq

\bibliography{refs}

\end{document}